\documentclass[preprint,showpacs,nofootinbib,prd,aps]{revtex4-1}

\usepackage{graphicx}
\usepackage{amstext}
\usepackage{amssymb}
\usepackage[usenames]{color}

\begin{document}
\title{Femtoscopy scales and particle production in the relativistic 
heavy ion collisions from Au+Au at 200 AGeV to Xe+Xe at 5.44 ATeV
within the integrated hydrokinetic model}

\author{V.~M.~Shapoval$^1$} 
\author{M.~D.~Adzhymambetov$^{1}$}
\author{Yu.~M.~Sinyukov$^{1}$}
\affiliation{$^1$Bogolyubov Institute for Theoretical Physics,
14b Metrolohichna street,  Kiev 03143, Ukraine}

\begin{abstract}
The recent results on the main soft observables, including hadron and photon yields 
and particle number ratios,  $p_T$ spectra, flow harmonics, as well as the 
femtoscopy radii, obtained within the integrated hydrokinetic model (iHKM) 
for high-energy heavy-ion collisions are reviewed and re-examined.
The cases of different nuclei colliding at different energies are 
considered: Au+Au collisions at the top RHIC energy $\sqrt{s_{NN}}=200$~GeV,
Pb+Pb collisions at the LHC energies $\sqrt{s_{NN}}=2.76$~TeV and 
$\sqrt{s_{NN}}=5.02$~TeV, and the LHC Xe+Xe collisions at $\sqrt{s_{NN}}=5.44$~TeV.
The effect of the initial conditions and the model parameters, including the 
utilized equation of state (EoS) for quark-gluon phase, on the simulation results,
as well as the role of the final afterburner stage of the matter evolution 
are discussed. The possible solution of the so-called ``photon puzzle'' is considered.
The attention is also paid to the dependency of the interferometry volume and 
individual interferometry radii on the initial transverse geometrical size of the 
system formed in the collision.
\end{abstract}

\maketitle


\section{Introduction}
Present-day the ultrarelativistic heavy-ion collision experiments, carried out 
at the BNL Relativistic Heavy Ion Collider (RHIC) and at the CERN Large Hadron
Collider (LHC), provide the only laboratory method of obtaining a new unusual 
quark-gluon state of matter, characterized by extremely high temperature and
energy densities. 
It is believed that this state is quite similar to the one the matter
had in the very early Universe at times about $10^{-6}$ seconds after the Big Bang. 
Naturally, the comprehensive study of the properties of matter under such extreme 
conditions and the dynamics of its evolution constitutes a fundamental physical
problem. 

In ultrarelativistic heavy-ion collision, in the energy range starting from top RHIC and higher,
due to strong Lorentz contraction the two colliding nuclei can be considered as 
ultrathin ``pancakes'' of quarks and gluons moving towards each other at a great 
speed~\footnote{The corresponding Lorentz $\gamma$ factor in the center of mass 
frame is of the order of $10^2$ for top RHIC energy and of the order of $10^3$ for the LHC. 
Thus, the thickness of the partonic ``pancake'' in each case is of the order 
of $10^{-2}$ fm and $10^{-3}$ fm respectively.}. 
After the nuclei pass through each other and carry away practically
all the net baryon charge, the space region between them becomes occupied by 
a hot and dense system of partons with small (for RHIC) or practically zero (for LHC) baryon chemical potential. Such a system fastly expands,
cools down, and eventually disintegrates into a system of about several thousand
hadrons and resonances (together with some amount of leptons, photons, 
electrons etc.). At this stage hadrons intensively interact with each other, 
experiencing elastic and inelastic scatterings, and the resonance decays take 
place. Finally, when all the interactions cease, produced free particles travel to 
the detectors, allowing the experimentalists to collect sets of data, from which one 
extracts various information about the created system. The entire complicated process of 
the system's evolution in each collision is extremely fast and takes only about $10^{-22}$ seconds. 

A thorough analysis of a great massive of measured data indicates that at the  
early stage of the collision the created strongly interacting matter gets 
thermalized and demonstrates the collective properties. And although the precise 
mechanism of the thermalization is still not clear (see \cite{therm1,therm2,
therm3,therm4,therm5,therm6,therm7,therm8,therm9} for the
corresponding discussion), the hypothesis that it takes 
place, and at some moment the system comes to a state of local thermal and chemical 
equilibrium, allows one to utilize the relativistic hydrodynamics 
approximation to describe the system's evolution during the time, while such 
equilibrium is preserved~\footnote{It is interesting, that although a new state of 
matter, created in relativistic nucleus-nucleus collisions is often referred 
to as ``quark-gluon plasma'' (QGP), to stress its property to contain 
free color charges, it actually behaves rather like a nearly perfect fluid, 
than like ``plasma'', i.e. like ionized gas.}.
The application of hydrodynamical formalism implies 
utilization of certain equation of state (EoS) for the considered hadron fluid, 
which cannot be strictly defined at the moment and thus is model dependent.

The representation of the system in terms of continuous medium is convenient, 
however, at the late times the matter loses the local thermal and chemical equilibrium,  the hydrodynamics 
becomes inapplicable, and the system transforms into a gas of particles 
(this transformation is usually considered to be connected with the 
restoration of QCD chiral symmetry and happening at the temperature close to 
150~MeV). Thus, for a proper description of the collision's final ``afterburner'' 
stage another type of a model is required, e.g. a hadron cascade model~\cite{urqmd1,urqmd2,jam}. 
So, now a theorist who wishes to develop an adequate approach for the relativistic 
heavy-ion collisions most likely will need to combine in the model different approaches for 
simulation of different stages of the matter evolution --- such collision models 
are known as hybrid ones~\cite{nonaka,hirano1,pratt,petersen,werner,song}.

In this paper we review the recent results on soft physics observables obtained 
for the high-energy heavy-ion collisions within the integrated hydrokinetic model 
(iHKM)~\cite{ihkm1,ihkm2}. The iHKM can be considered as hybrid model as well, since 
it includes viscous hydrodynamics for description of locally equilibrated phase and
switching to the UrQMD hadron cascade~\cite{urqmd1,urqmd2} at the particlization isothermal 
hypersurface $T_\mathrm{p}$. However, the advantage of iHKM is that it includes treatment of 
pre-equilibrium dynamics of the system, which most other models lack.

Typically the collision simulation in a hybrid model starts right from the 
hydrodynamics at relatively large times, $\tau \sim 0.6-1$~fm/$c$, when 
the system can be expected to become nearly thermalized. However, the initial state of
the matter is associated with a certain model energy (or entropy) density 
distribution, related to a very early evolution moment just after 
the collision of the two nuclei (e.g. the Monte Carlo Glauber \texttt{GLISSANDO} model~\cite{gliss} 
or the Monte Carlo Kharzeev-Levin-Nardi model~\cite{kln}). 
Such states are typically characterized by large anisotropy in the local 
rest frame momentum spaces and are far from equilibrium~\cite{gelis}. It means that they
cannot serve as proper initial conditions for hydrodynamics.
Some equilibration process, transforming this primordial distribution to 
a nearly hydrodynamical form, must precede the system expansion described in terms of a continuous fluid.

In contrast to other models, in iHKM the simulation starts at the early time $\tau_0 \sim 0.1$~fm/$c$
with the pre-thermal dynamics stage, during which
a far-from-equilibrium energy-momentum tensor of the system gradually evolves 
to a relativistic viscous hydrodynamics locally (partially) equilibrated tensor, which
in turn serves as the initial condition for the hydrodynamical stage.
Such an early beginning of the matter evolution is important for successful
description of data, since the collective effects develop noticeably just because of 
finiteness and azimuthal asymmetry of the system already at the pre-equilibrium 
stage without any pressure gradients~\cite{ihkm1}. 

Another known issue related to hybrid models concerns possible violations of
the energy-momentum conservation law, when one switches from continuous medium
to particle gas expansion at the hypersurface, which contains non-spacelike 
elements, using Cooper-Frye prescription.
The hydrokinetic approach, previously developed in Ref.~\cite{hkm1,hkm21,hkm22,hkm23,hkm3} (see 
also Ref.~\cite{hkm4}) and implemented in the hydrokinetic model (HKM) --- the 
predecessor of iHKM --- allows to avoid this problem and consider the realistic 
continuous emission of particles from the system all along the process of its 
hydrodynamical evolution through the use of escape function formalism instead of the 
distribution function one. However, in~\cite{hkm3} it was also shown that if one
uses smooth initial conditions, obtained by averaging of a large set of single-event 
density profiles, the contribution from non-spacelike parts of the hadronization
hypersurface will be less than 2\% and can be neglected without bringing noticeable 
distortion to the final simulation results (this can be explained by small
spatial sizes of the corresponding hypersurface elements at small $\tau$
and strong collective flows at large $\tau$). For this reason, and because of
rather time-consuming calculations required for a full-value hydrokinetic approach 
realization, we used sudden switching to hadronic cascade at the particlization 
isotherm in the iHKM studies discussed here.

In the recent papers \cite{ihkm2,ratiosour,lhc502-ihkm,rhic-ihkm} a simultaneous description
and prediction of various bulk observables including the femtoscopy radii 
were obtained in the iHKM for the LHC Pb+Pb collisions at $\sqrt{s_{NN}}=2.76$~TeV 
and $\sqrt{s_{NN}}=5.02$~TeV, as well as for Au+Au collisions at the top RHIC 
energy $\sqrt{s_{NN}}=200$~GeV, using a single set of model parameters for each 
collision type. In Ref.~\cite{ratiosour} the role of the post-hydrodynamic stage of
the system's evolution in particle production was investigated. 
In~\cite{kstar} the problem of resonances observability in view
of the interaction of their decay products with the hadronic medium created at
the late stage of the collision was analyzed within iHKM.
The model allowed also to describe photon production
(momentum spectra and $v_2$ coefficients) for the cases of top RHIC energy and  
LHC energy $2,76 A$~TeV~\cite{photons1,photons2}. As a result, an all-around 
detailed picture of the matter evolution and particle emission in course of a 
relativistic heavy-ion collision was obtained, and the space-time structure of the 
created system was revealed.
Here we summarize the works~\cite{ihkm2,ratiosour,lhc502-ihkm,rhic-ihkm,kstar}
concerning hadronic observables, direct photon production \cite{photons1,photons2} and present a newly obtained 
results for Xe+Xe collisions at the LHC energy $\sqrt{s_{NN}}=5.44$~TeV.

\section{The model description}
The full complicated evolution process of the system, created in ultrarelativistic
heavy-ion collision, can be divided into several stages. Each stage description
within iHKM is realized using an appropriate formalism.

\subsection{Initial pre-equilibrium state formation} 
This stage models the earliest
far-from-equilibrated energy-momentum distribution of partonic system, formed
just after the nuclei collision. It is attributed to the initial proper time 
$\tau_0 \sim 0.1$~fm/$c$ and serves as a starting point for the subsequent
pre-thermal relaxation energy-momentum transport dynamics, which gradually 
transforms the non-equilibrated energy-momentum tensor to the one describing 
a thermalized state of the system in nearly local equilibrium. 
The latter should be reached by the thermalization time, believed to be close to 
the inverse pion mass, $\tau_\mathrm{th} \approx 1$~fm/$c$. We assume the initial 
parton distribution function to have a factorized form 
\begin{equation}
\label{fdist}
f(t_{\tau_0},\textbf{r},p)=\epsilon(b,\tau_0,\textbf{r}_{T})f_0(\eta,p).
\end{equation}
The transverse energy-density profile $\epsilon(b,\tau_0,\textbf{r}_{T})$
is generated using the \texttt{GLISSANDO} code~\cite{gliss}, 
that implements Monte Carlo Glauber approach with parameter $\alpha$ that defines contribution to energy density from binary collisions ({\it bin}), correspondingly $1-\alpha$ is related to contribution from wounded ({\it w}) nucleons:
\begin{equation}
\label{eps0}
\epsilon(b,\tau_0,\textbf{r}_{T})=\epsilon_{0}(\tau_0)
\frac{(1-\alpha)N_w(b,\textbf{r}_{T})/2+\alpha N_{bin}(b,\textbf{r}_{T})}{(1-\alpha)N_w(b=0,\textbf{r}_{T}=0)/2+\alpha N_{bin}(b=0,\textbf{r}_{T}=0)}.
\end{equation}
The boost-invariant momentum function $f_0(\eta,p)$,  where  $\eta=\tanh^{-1}(z/t)$ is the space-time rapidity, corresponds to the Color Glass Condensate (CGC) effective gluon field theory:
\begin{equation}
f_0(p)=g \exp\left(-\sqrt{\frac{(p\cdot U)^2-(p\cdot V)^2}{\lambda_{\perp}^2}+
\frac{(p\cdot V)^2}{\lambda_{\parallel}^2}}\right),
\label{aniso}
\end{equation}
where $U^{\mu}=( \cosh\eta, 0, 0, \sinh\eta)$, $
V^{\mu}=(\sinh\eta, 0, 0,\cosh\eta)$.
The parameters $\lambda_{\parallel}$ and $\lambda_{\perp}$ in (\ref{aniso}) 
can be interpreted
as the temperatures parallel and perpendicular to the beam axis, since
in the  ``rest frame'' $\eta=0$, $(p\cdot U)^2-(p\cdot V)^2=p_{\perp}^2$
and $(p\cdot V)^2=p_{\parallel}^2$. Then one can introduce 
a model parameter $\Lambda=\lambda_{\perp}/\lambda_{\parallel}$, defining the 
initial state momentum anisotropy. Its value is fixed at $\Lambda=100$ for all
the considered experimental set-ups. Such a large anisotropy of the initial state
is specific for the models based on the Color Glass Condensate theory. 

\begin{figure}
\centering
\includegraphics[bb=0 0 567 383, width=0.49\textwidth]{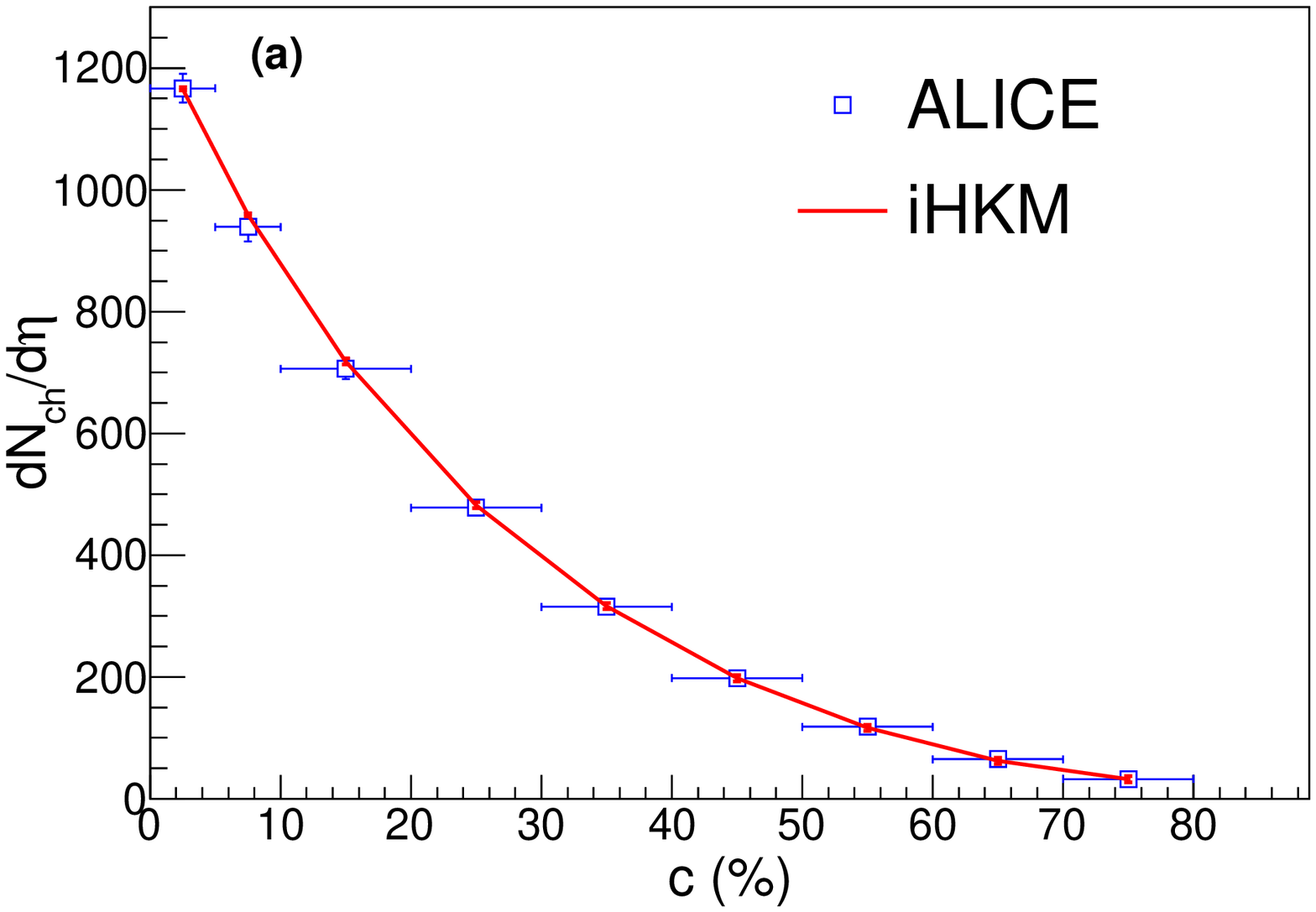}
\includegraphics[bb=0 0 567 409, width=0.49\textwidth]{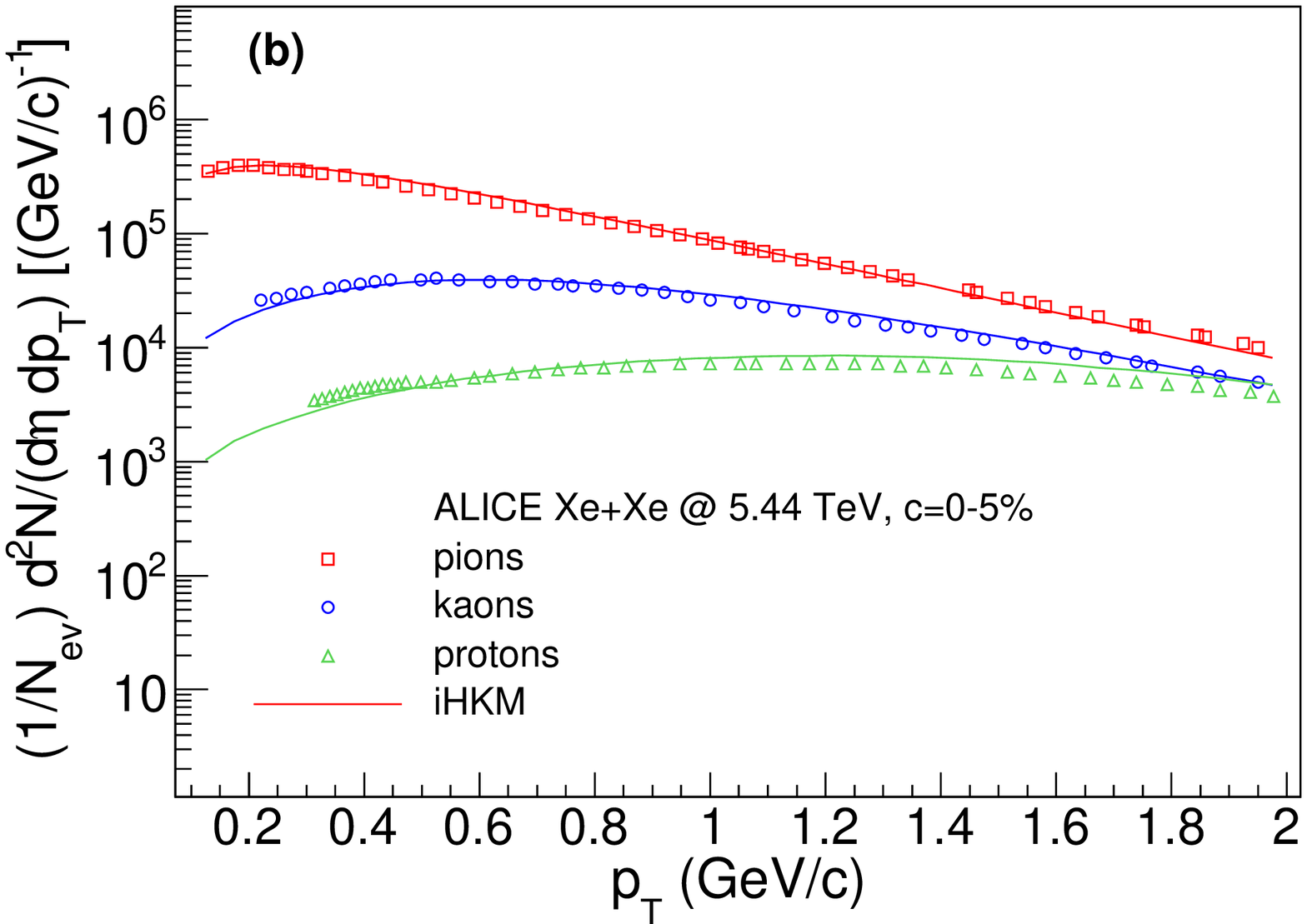}
\caption{The calibration of the iHKM parameters is performed based on the 
mean charged particle density $\langle dN_{\mathrm{ch}}/d\eta \rangle$ 
dependency on centrality (a) and the $p_T$ spectra slope 
in the most central events~(b). The best combined fit within the model 
defines for a given collision type the corresponding values of $\alpha$ and 
$\epsilon_0(\tau_0)$ parameters. In this figure the data and model curves
for the LHC Xe+Xe collisions at $\sqrt{s_{NN}}=5.44$~TeV are shown.
The ALICE Collaboration experimental points are taken 
from~\cite{alicexe,alicexe2}.
\label{mult1}} 
\end{figure}

The energy density distribution $\epsilon(b,\textbf{r}_{T})$ in~(\ref{eps0})
includes two contributions: $N_{bin}(b,\textbf{r}_T)$ from the binary collisions 
and $N_w(b,\textbf{r}_T)$ from the wounded nucleons models, where $b$ is the 
impact parameter denoting the collision centrality. The parameter~$\alpha$ 
regulates the proportion between these two terms, 
and $\epsilon_0(\tau_0)$ defines the maximal energy density in the center of the 
system at the initial time $\tau_0$ in the most central events. 
The values of $\alpha$ and $\epsilon_0(\tau_0)$ 
are fixed based on experimental mean charged particle density dependence on 
centrality and the pion $p_T$ spectrum slope in central events 
(see Fig.~\ref{mult1}) and serve as the two main iHKM parameters that adjust 
the model for the description of a particular collision type. These parameters 
have the same values for all the centrality classes, while the collision 
centrality is regulated by the \texttt{GLISSANDO} options, specifying cuts on the 
participant nucleons number.

Note also, that while \texttt{GLISSANDO} allows one to generate density profiles
of single collision events, which can be then used for an event-by-event analysis,
in our current studies we used a single smooth initial state profile, obtained by 
averaging of 50000 \texttt{GLISSANDO} events, for simulation of each 
particular collision type at given centrality within iHKM.

\subsection{Pre-thermal matter evolution} 
During this stage the non-equilibrated 
initial energy-momentum tensor $T^{\mu \nu}_{\mathrm{0}}(x)$ smoothly evolves, 
being little by little ``mixed'' with the Israel-Stewart tensor 
$T_{\mathrm{hydro}}^{\mu \nu}(x)$, corresponding to viscous hydrodynamics regime. 
We use the relaxation time approximation associated with the equations of 
energy-momentum conservation for the tensor $T^{\mu \nu}(x)$ to evaluate the latter 
in course of its pre-equilibrium evolution. An important advantage of the applied 
formalism is that it does not require any additional assumptions, like the 
``anisotropic equilibrium''~\cite{anis1,anis2,anis3} or the Landau matching conditions, to 
continuously transfer the initial anisotropic and non-thermalized state at $\tau_0$ 
to the nearly thermalized one at $\tau_\mathrm{th}$. Also, the utilized method is suitable 
for event-by-event simulations, since it allows to take into account large 
inhomogeneities of the system initial state, which will likely result in unusual 
transverse dynamics.

In the relaxation time approximation for Boltzmann equation
the energy-momentum tensor of expanding matter $T^{\mu \nu}(x)$ 
at arbitrary moment of time between $\tau_0$ and $\tau_\mathrm{th}$ can be written as 
follows~\cite{AkkSin} 
\begin{equation}
T^{\mu \nu}(x)=T^{\mu \nu}_{\mathrm{free}}(x){\cal P}(\tau)+T_{\mathrm{hydro}}^{\mu \nu}(x)[1-{\cal P}(\tau)].
\label{tbase}
\end{equation}
Here $T^{\mu \nu}_{\mathrm{free}}(x)$ is an energy-momentum tensor, 
constructed based on the distribution function 
$f_{\mathrm{free}}(t,\textbf{r},\textbf{p})=f(t_{\tau_0},\textbf{r}-\frac{\textbf{p}}{p_0}(t-t_{\tau_0}),\textbf{p})$,
which corresponds to a nearly free-streaming evolution of the initial
non-equilibrium distribution (\ref{fdist}), i.e.
\begin{equation}
T^{\mu \nu}_{\mathrm{free}}(x)=\int d^{3}p \frac{p^{\mu}p^{\nu}}{p_{0}}f_{\mathrm{free}}(x,p). 
\label{tmn1}
\end{equation}
The tensor $T_{\mathrm{hydro}}^{\mu \nu}(x)$ is defined by the Israel-Stewart 
relativistic viscous hydrodynamics formula: 
\begin{equation}
T^{\mu \nu}_{\mathrm{hydro}}(x)= [\epsilon_{\mathrm{hydro}}(x) +
p_{\mathrm{hydro}}(x)+\Pi]u^{\mu}_{\mathrm{hydro}}(x)u^{\nu}_{\mathrm{hydro}}(x)
- [p_{\mathrm{hydro}}(x)+\Pi]g^{\mu \nu} +\pi^{\mu \nu},
\label{tmn2}
\end{equation}
where $\epsilon_{\mathrm{hydro}}$ is the local rest frame energy density, 
$p_{\mathrm{hydro}}$ is the local rest frame pressure, $\pi^{\mu \nu}$ is the shear 
stress tensor, $u^{\mu}_{\mathrm{hydro}}(x)$ is the four-vector of energy flow, 
$g^{\mu \nu}$ is the metric tensor and $\Pi$ is the bulk pressure. 
In our current studies we neglect the bulk pressure term, and so put $\Pi=0$. 
The shear stress tensor, being independent dynamical variable, requires
its own separate equation of motion, which we write, neglecting the vorticity
terms, as follows:
\begin{equation}
\langle u^\gamma \partial_{;\gamma} \pi^{\mu\nu}\rangle
=-\frac{\pi^{\mu\nu}-\pi_\mathrm{NS}^{\mu\nu}}{\tau_\pi}-\frac 4 3
\pi^{\mu\nu}\partial_{;\gamma}u^\gamma. 
\label{mov1}
\end{equation}
In this equation the semicolon denotes a covariant derivative,
$\pi^{\mu\nu}_\mathrm{NS}$ is the Navier-Stokes shear stress tensor,
$\pi^{\mu\nu}_\mathrm{NS}=\eta(\Delta^{\mu\lambda}\partial_{;\lambda}u^\nu+\Delta^{\nu\lambda}\partial_{;\lambda}u^\mu)-\frac 2 3 \eta\Delta^{\mu\nu}\partial_{;\lambda}u^\lambda$, and brackets mean the following operation: 
\begin{equation}
\langle A^{\mu\nu}\rangle=\left( \frac 1 2 \Delta^\mu_\alpha
\Delta^\nu_\beta+\frac 1 2 \Delta^\nu_\alpha \Delta^\mu_\beta -
\frac 1 3 \Delta^{\mu\nu}\Delta_{\alpha\beta}\right)A^{\alpha\beta}, 
\end{equation}
where $\Delta^{\mu\nu}=g^{\mu\nu}-u^\mu u^\nu$.

The function ${\cal P}(\tau)$ in Eq.~(\ref{tbase}) is the weight function,
which has to satisfy the following conditions:
\begin{equation}
0\leq{\cal P}(\tau)\leq1, \quad {\cal P}(\tau_0)=1, \quad {\cal P}(\tau_\mathrm{th})=0, \quad \mathrm{and} \quad \partial_{\mu}{\cal P}(\tau_\mathrm{th})=0. 
\label{constr}
\end{equation}
Within the Boltzmann relaxation kinetics formalism, one can express the weight
function in the following general form~\cite{AkkSin}:
\begin{equation}
{\cal P}(\tau,\textbf{r},p)=\exp{\left(-\int_{\tau_0}^{\tau} 1/\tau_\mathrm{rel}(\tau',\textbf{r},p)d\tau'\right)}.
\end{equation}
If one assumes that the relaxation time $\tau_\mathrm{rel}(x,p)$ here depends 
only on $\tau$ for each fluid element, then ${\cal P}(\tau,\textbf{r},p)$
also will depend only on $\tau$.
This consideration corresponds to the Bjorken picture~\cite{bjorken}, 
assuming the thermalization of the matter to develop synchronously in  
proper time of different fluid elements. Then one will have:
\begin{equation}
{\cal P}(\tau)=\exp{\left(-\int_{\tau_0}^{\tau} \frac{1}{\tau_\mathrm{rel}(\tau')}d\tau'\right)}.
\label{calp}
\end{equation}
Taking $\tau_\mathrm{rel}(\tau')$ in the form $\tau_\mathrm{rel}(\tau')=\tau_\mathrm{rel}(\tau_0)\frac{\tau_\mathrm{th}-\tau'}{\tau_\mathrm{th}-\tau_0}$, 
one can easily perform the integration in~(\ref{calp}), and obtain the expression
for ${\cal P}(\tau)$:
\begin{equation}
{\cal P}(\tau)= \left(\frac{\tau_\mathrm{th}-\tau}{\tau_\mathrm{th}-\tau_0}\right)^{\frac{\tau_\mathrm{th}-\tau_0}{\tau_\mathrm{rel}(\tau_0)}}. 
\end{equation}
Here $\tau_\mathrm{rel}(\tau_0)\equiv\tau_\mathrm{rel}$ is one of the iHKM parameters. 
In accordance with Eq.~(\ref{constr}), it is required that 
$(\tau_\mathrm{th}-\tau_0)/\tau_\mathrm{rel}>1$.

The evolution of the system's energy-momentum tensor will be governed 
by the relativistic hydrodynamics equations
\begin{equation}
\partial_{;\mu}\{[1-{\cal P}(\tau)]T^{\mu\nu}_{\mathrm{hydro}}(x)\}= 
- T^{\mu\nu}_{\mathrm{free}}(x)\partial_{;\mu}{\cal P}(\tau),
\label{evol}
\end{equation}
following from the energy-momentum conservation law for the full energy-momentum
tensor, $\partial_{;\mu}T^{\mu\nu}(x)=0$.
One can see, that (\ref{evol}) can be re-written as the hydrodynamics equations 
for a new energy-momentum tensor 
$\tilde{T}^{\mu\nu}_{\mathrm{hydro}}(x)=[1-{\cal P}(\tau)]T^{\mu\nu}_{\mathrm{hydro}}(x)$, such that $\widetilde{T}^{\mu \nu}_{\mathrm{hydro}}(x)=0$ 
at $\tau = \tau_{0}$ for all $x$, 
and with an additional source term on the right-hand side:  
\begin{equation}
\partial_{;\mu}\widetilde{T}^{\mu \nu}_{\mathrm{hydro}}(x)= - T^{\mu \nu}_{\mathrm{free}}(x)\partial_{;\mu}{\cal P}(\tau). 
\label{mov3}
\end{equation}
The shear stress tensor also has to be re-defined as 
$\tilde{\pi}^{\mu\nu}=[1-{\cal P}(\tau)]\pi^{\mu\nu}$, so that the corresponding 
equation of motion (\ref{mov1}) will take the form:
\begin{equation}
[1-{\cal P}(\tau)]\left \langle u^\gamma \partial_{;\gamma}
\frac{\widetilde{\pi}^{\mu\nu}}{(1-{\cal P}(\tau))}\right \rangle
=-\frac{\widetilde{\pi}^{\mu\nu}-[1-{\cal
P}(\tau)]\pi_\text{NS}^{\mu\nu}}{\tau_\pi}-\frac {4}{3}
\widetilde{\pi}^{\mu\nu}\partial_{;\gamma}u^\gamma.
\label{mov4}
\end{equation}
Since $T^{\mu \nu}_{\mathrm{free}}(x)$ and ${\cal P}(\tau)$ are known, 
one can numerically find the solutions of the above equations with respect to 
$\widetilde{T}^{\mu \nu}_{\mathrm{hydro}}(x)$, and thus describe
the pre-equilibrium dynamics of the system and gradual transition of the tensor
$\widetilde{T}^{\mu \nu}_{\mathrm{hydro}}(x)$ to hydrodynamical regime
by the time of thermalization,
$\widetilde{T}^{\mu \nu}_{\mathrm{hydro}}(x) \rightarrow T^{\mu \nu}_{\mathrm{hydro}}(x)$ 
when $\tau \rightarrow \tau_\mathrm{th}$.
However, to obtain such a description one also needs to close the system of 
hydrodynamical equations, which define the matter evolution, and specify the 
equation of state for considered quark-gluon fluid.

\subsection{Hydrodynamics stage}
At the hydrodynamical stage, which starts at $\tau=\tau_\mathrm{th}$, the evolution
of the system is described in terms of hadron fluid expansion within the
framework of Israel-Stewart relativistic viscous hydrodynamics. 
The ratio of shear viscosity to entropy density parameter is fixed to a value 
$\eta/s = 0.08 \approx 1/4\pi$, i.e. close to its minimum possible value,
according to Ref.~\cite{ihkm2}. The viscous hydrodynamics equations (\ref{mov3})
and (\ref{mov4}) are numerically solved in iHKM using the \texttt{vHLLE} code~\cite{vhlle}.

The hydrodynamical evolution of the system continues until it loses
local thermal and chemical equilibrium and transforms into hadron-resonance gas.
It happens when the local temperature becomes lower than the particlization
temperature $T_\mathrm{p}$. The value of this temperature depends on the utilized
equation of state for quark-gluon matter. In our studies we used the two
QCD-inspired equations of state: Laine-Schroeder EoS~\cite{EoS} with the 
corresponding particlization temperature $T_\mathrm{p}=165$~MeV and HotQCD
Collaboration EoS~\cite{EoS2} with $T_\mathrm{p}=156$~MeV, consistent
with the latest thermal model estimates as for the temperature of chemical 
freeze-out, $T_\mathrm{ch}=156\pm1.5$~MeV~\cite{stachel-sqm2013}. 
Also, both particlization temperatures are close to the Lattice QCD estimate 
for pseudo-critical temperature in the cross-over scenario, 
$T_\mathrm{ps}=154\pm9$~MeV.

In case of the RHIC energies, unlike in the LHC case, the utilized EoS needs to be corrected for a 
non-zero baryon and strange chemical potentials~\cite{eos-cor}:
\begin{equation}
\label{eos-corr}
\frac{p(T,\mu_B,\mu_S)}{T^4}=\frac{p(T,0,0)}{T^4}+\frac{1}{2}\frac{\chi_B}{T^2}{\left( \frac{\mu_B}{T}\right)}^2 +\frac{1}{2}\frac{\chi_S}{T^2}{\left( \frac{\mu_S}{T}\right) }^2,
\end{equation}
where $p(T,0,0)$ is the uncorrected equation of state at zero chemical potentials and
\begin{equation}
\frac{\chi_i}{T^2}=\frac{1}{VT^3}\frac{\partial^2 \mathrm{ln}Z}{\partial(\mu_i/T)^2}, \ \ \  i=B,S.
\end{equation}
In our studies, devoted to the Au+Au collisions at RHIC~\cite{rhic-ihkm} we put
$\mu_B=21$~MeV at the particlization isotherm $T_\mathrm{p}=165$~MeV, ensuring
the best description of the proton to antiproton yields ratio in the central events with $c=0-5\%$. 
The strange chemical potential value $\mu_S=5$~MeV was fixed based on the  
condition of zero strangeness at the particlization hypersurface:
\begin{equation}
\left. S\right|_{\sigma_{p}}=\sum_{i}[N(i)-\overline{N}(i)]\mu_{S,i}=0, 
\end{equation}
where the summation is done over all the particle sorts, $N(i)$ and $\overline{N}(i)$ are the numbers 
of particles and antiparticles of the $i^\mathrm{th}$ species at the hypersurface $T_\mathrm{p}$, 
and $\mu_{S,i}$ is the particle strange chemical potential. 

\subsection{Particlization stage}
At the particlization stage we switch from the continuous medium to particles 
cascade in the matter evolution description, using the common 
Cooper-Frye prescription,
\begin{equation}
p^0 \left.\frac{d^3 N_i(x)}{d^3 p}\right|_{d\sigma(x)} = 
d\sigma_{\mu}(x) p^{\mu} f_i(p \cdot u(x), T(x), \mu_i(x)).
\label{CF1}
\end{equation}
The subscript $i$ here denotes the particle sort number and $d\sigma(x)$ 
is the element of the switching hypersurface $\sigma_{\mathrm{sw}}$, related to the
space-time point $x$.
To build the particlization hypersurface we utilize the Cornelius 
routine~\cite{cornelius1,cornelius2,cornelius3}. 
The functions $f_i(x,p)$ in (\ref{CF1}) are obtained
after application of the Grad ansatz~\cite{grad} viscous corrections 
to the corresponding local equilibrium distribution functions $f^{\mathrm{l.eq.}}_i(x,p)$. 
Assuming the same corrections for all hadron types, one can write (\ref{CF1})
as follows: 
\begin{equation}
\frac{d^3\Delta N_i}{dp^{*}d(\cos{\theta})d\phi}=\frac{\Delta \sigma^{*}_{\mu}p^{*\mu}}{p^{*0}}p^{*2} f^{\mathrm{l.eq.}}_{i}\left(p^{*0};T,\mu_i\right)\left[1+(1\mp f^{\mathrm{l.eq.}}_{i})\frac{p^*_\mu p^*_\nu \pi^{*\mu \nu}}{2T^2(\epsilon+p)}\right].
\label{CF2}
\end{equation}
Using this distribution, one can calculate the average particle number $N_i$
for each species~$i$. The sum of all these numbers will give the average total
number of particles $N_{\mathrm{tot}}$ at the particlization 
hypersurface~$\sigma_{\mathrm{sw}}$.
Then one can use Monte Carlo method to generate particles with their 
momenta and coordinates in accordance with (\ref{CF2}).
The total number of particles in a given event obeys the Poisson distribution
with the mean value $N_{\mathrm{tot}}$, and the species of each particle is 
selected randomly with probability $N_i/N_{\mathrm{tot}}$.

\subsection{Afterburner stage}
At the final post-hydrodynamical stage of the system's evolution all the 
particles, generated at the switching hypersurface are passed to
the UrQMD hadron cascade code~\cite{urqmd1,urqmd2}. Since in iHKM we aim to account for
all reliably known hadron resonance states, while many of them are not processed by 
UrQMD, the heavy resonances not present in the UrQMD particle database, 
are decayed right at the particlization isotherm $T_\mathrm{p}$ to ensure 
the energy-momentum conservation.

The UrQMD cascade code performs elastic and inelastic scatterings between hadrons
and decays the particles known to be unstable. The program has multiple options,
which allow one to change tuning of the model, e.g. switch baryon-antibaryon 
annihilation on/off or disallow some resonance decays to adjust the simulation 
to the experimental treatment of secondary particles.

As our previous studies demonstrate~\cite{kstar,ratiosour,lhc502-ihkm}, the 
afterburner stage plays an important role in the formation of particle yields,
their ratios and spectra. This supports the hypothesis of continuous (not sharp) 
character of chemical and thermal freeze-out in heavy-ion collisions.

\section{Results and discussion}

\subsection{Model calibration}
To start a certain type of collision simulation in iHKM, one firstly needs to 
calibrate the model, setting the fitting values of its parameters, that would
result in a proper description of the system's evolution process.

As it was shown in \cite{ihkm2,lhc502-ihkm}, the main time parameter, 
strongly affecting the formation of observables is the initial time $\tau_0$,
while the thermalization time $\tau_\mathrm{th}$, defining the moment of switching
from pre-thermal dynamics to hydrodynamics, when the system comes to a nearly
thermalized state, and the relaxation time $\tau_\mathrm{rel}$, defining the rate
of thermalization process, can be varied within quite wide limits and still
give a good description of measured data, if the initial maximal energy-density
parameter $\epsilon_0(\tau_0)$ is accordingly re-adjusted. 
Thus, for all the collision set-ups considered here the same setting of 
$\tau_\mathrm{rel}$ and $\tau_\mathrm{th}$ are used, namely 
$\tau_\mathrm{rel}=0.25$~fm/$c$ and $\tau_\mathrm{th}=1$~fm/$c$.
The same concerns the initial state momentum anisotropy parameter, $\Lambda$.
As it was mentioned above, its value is fixed to $\Lambda=100$ for all types
of collisions under consideration.

The parameter $\alpha$, defining the fraction of binary collision model contribution
to the initial energy density distribution (\ref{eps0}), has the smallest
value in iHKM for Au+Au collisions at the top RHIC energy, $\alpha=0.18$,
a higher value $\alpha=0.24$ for Pb+Pb collisions at both $2.76 A$~TeV and
$5.02 A$~TeV energies and the maximum value $\alpha=0.44$ for Xe+Xe collisions
at $5.44 A$~TeV. The increasing of $\alpha$ parameter could be connected with
the increase of the inelastic nucleon-nucleon cross-section from $\sigma=42$~mb
in the RHIC Au+Au collisions case to $\sigma=64$~mb for the LHC Pb+Pb collisions
and $\sigma=68.4\pm5.0$~mb for the LHC Xe+Xe collisions. 

Apart from that, gold and lead nuclei have close radius values, about 6.37~fm and 
6.62~fm respectively, while xenon nucleus has noticeably smaller radius about 
5.36~fm. This also can lead to increasing of $\alpha$ value when going from Au+Au 
and Pb+Pb to Xe+Xe collisions, since the binary collisions model produces more 
narrow density distribution, which should be more appropriate in case of collision
of smaller nuclei.
Note also, that constructing the initial state for a Xe+Xe collision,
one should account for a slightly prolate shape of xenon nuclei, characterized
by the deformation parameter $\beta=0.18\pm0.02$~\cite{alicexe}. Such an accounting
can be realized using native \texttt{GLISSANDO} settings, so in our simulations we
just put the values of corresponding \texttt{GLISSANDO} parameters \texttt{BETA2A}
and \texttt{BETA2B} to 0.18.

The parameter $\tau_0$ defines the initial time at which the pre-equilibrium
stage of the matter evolution in iHKM starts. This time is associated with a 
far-from-equilibrium system state, formed very soon after the two nuclei 
overlapping in course of their collision. The estimates made in~\cite{lappi}
suggest the values $\tau_0 \approx 0.1$~fm/$c$ for the top RHIC energy collisions
and $\tau_0 \approx 0.07$~fm/$c$ for the full LHC energy case.
In our works we primarily used $\tau_0=0.1$~fm/$c$ for all the considered 
colliding systems. For the description of the top RHIC energy Au+Au collisions
the Laine-Schroeder equation of state for quark-gluon phase and the particlization 
temperature $T_\mathrm{p}=165$~MeV were used, while for the case of the LHC Xe+Xe 
collisions at $\sqrt{s_{NN}}=5.44$~TeV a more recent HotQCD Collaboration EoS with 
$T_\mathrm{p}=156$~MeV was chosen. 
In case of LHC Pb+Pb collisions the simulations were carried out using 
both mentioned equations of state with their corresponding particlization 
temperatures. Here the basic value $\tau_0=0.1$~fm/$c$ was used with 
the Laine-Schroeder EoS, and in case of the HotQCD EoS very close starting 
time values $\tau_0=0.15$~fm/$c$ for $\sqrt{s_{NN}}=2.76$~TeV and 
$\tau_0=0.12$~fm/$c$ for $\sqrt{s_{NN}}=5.02$~TeV were applied in order to reach 
the best fitting of pion $p_T$ spectrum at the model calibration.
The corresponding values of the initial maximal energy density parameter
$\epsilon_0(\tau_0)$ and the fraction of the binary collisions contribution
to the initial energy-density profile $\alpha$ are listed in the Table~\ref{tab1}.

\begin{table}
\centering
\begin{tabular}{|c|c|c|c|c|}
 \hline
 Experiment & EoS & $\alpha$ & $\tau_0$ (fm/$c$) & $\epsilon_{0}$ (GeV/fm$^3$)  \\ 
 \hline
 Au+Au @ 200 AGeV  & L.-S.  & 0.18 & 0.1 & 235    \\
 Pb+Pb @ 2.76 ATeV & L.-S.  & 0.24 & 0.1 & 679    \\
 Pb+Pb @ 2.76 ATeV & HotQCD & 0.24 & 0.15 & 495    \\
 Pb+Pb @ 5.02 ATeV & L.-S.  & 0.24 & 0.1 & 1067   \\
 Pb+Pb @ 5.02 ATeV & HotQCD & 0.24 & 0.12 & 870   \\
 Xe+Xe @ 5.44 ATeV & HotQCD & 0.44 & 0.1 & 445    \\
\hline
\end{tabular}
\caption{The iHKM parameters $\alpha$ and $\epsilon_{0}(\tau_0)$, used to tune
the model to the description of the listed heavy-ion collision events. } 
\label{tab1}
\end{table}

\subsection{Quark-gluon equations of state and freeze-out problem}
The comparison of the simulation results obtained using different quark-gluon
fluid equations of state can help to clarify the interplay between the
factors, defining the behaviour of experimentally measured observables at
hydrodynamical and post-hydrodynamical stages of the matter evolution.
Note, that the applied equation of state here should be considered rather as
the effective one, since the expansion rates of the system, formed in the
ultrarelativistic heavy ion collision, are extremely high (even much higher,
than in the Early Universe), so that the considered process is essentially
dynamical (non-static) and thus can hardly be adequately described by the
EoS, based on the lattice QCD calculations for the static system.

\begin{figure}
\centering
\includegraphics[bb=0 0 567 409,width=0.49\textwidth]{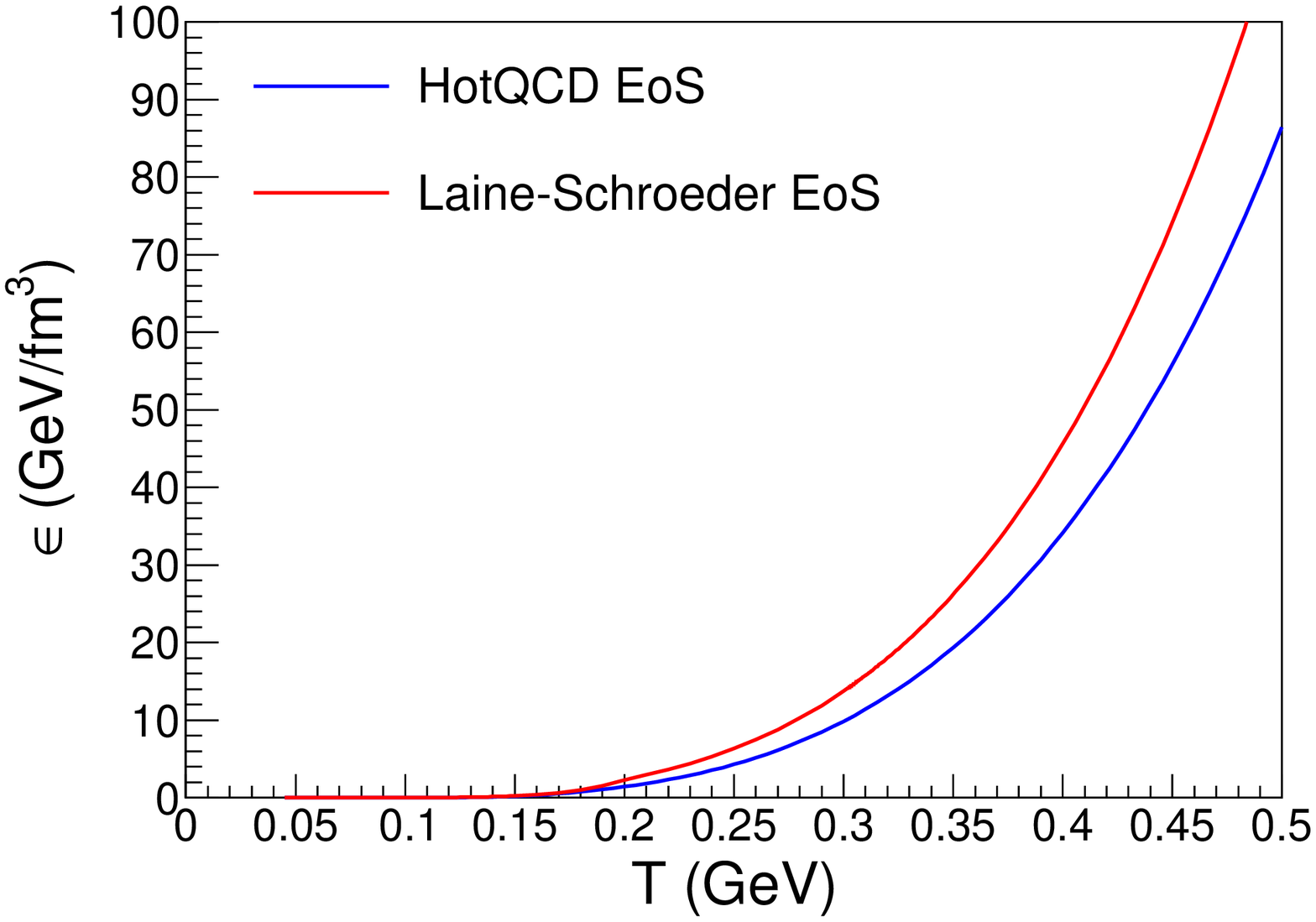}
\includegraphics[bb=0 0 567 409,width=0.49\textwidth]{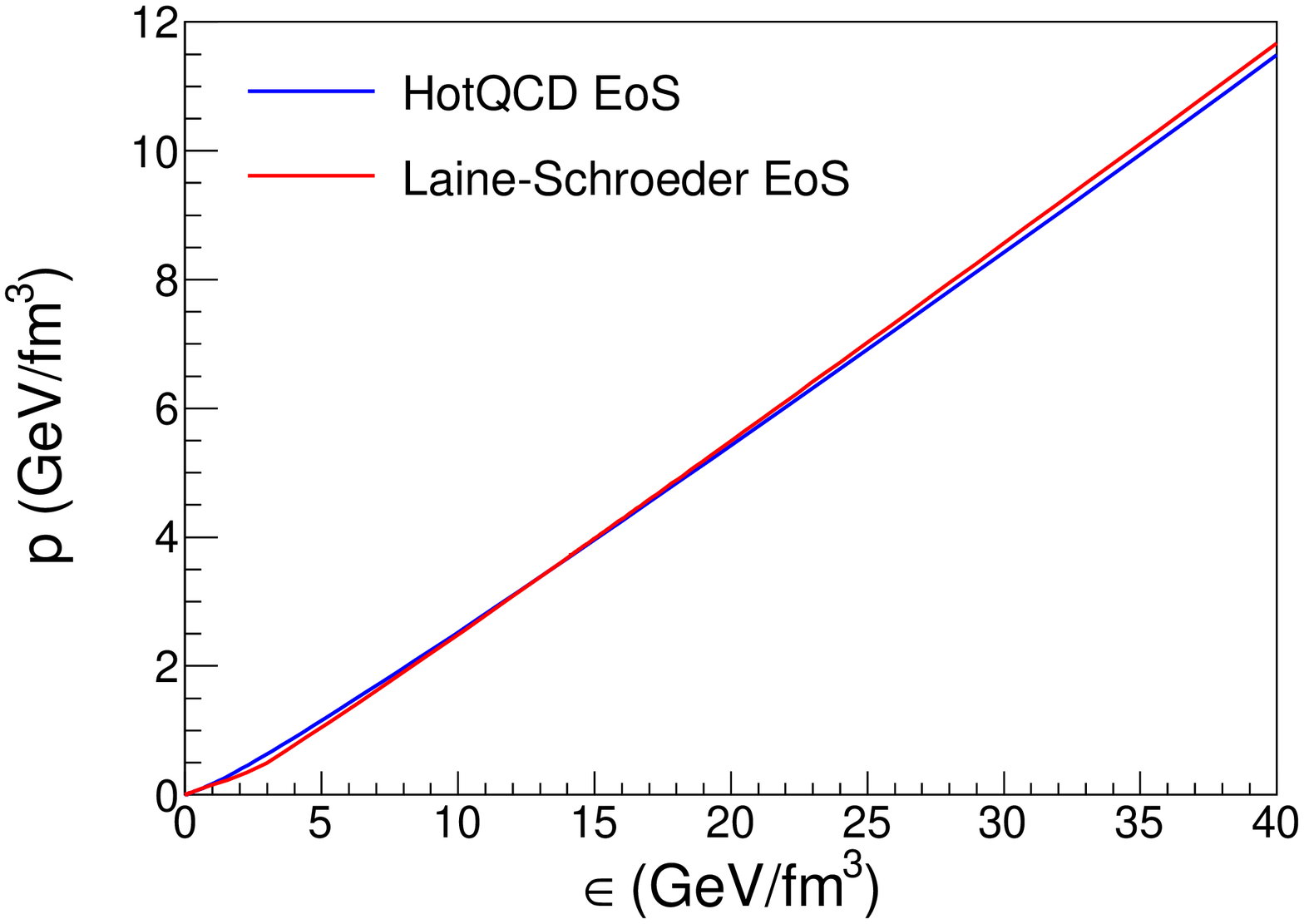}
\caption{The comparison of the two quark-gluon phase equations of state,
applied in iHKM simulations. 
The red line is related to the Laine-Schroeder EoS~\cite{EoS} with the 
temperature of particlization $T_\mathrm{p}=165$~MeV, and the blue line 
corresponds to the HotQCD Collaboration EoS~\cite{EoS2} with 
$T_\mathrm{p}=156$~MeV.
\label{eos2}}
\end{figure}

\begin{figure}
\centering
\includegraphics[bb=0 0 567 409,width=0.49\textwidth]{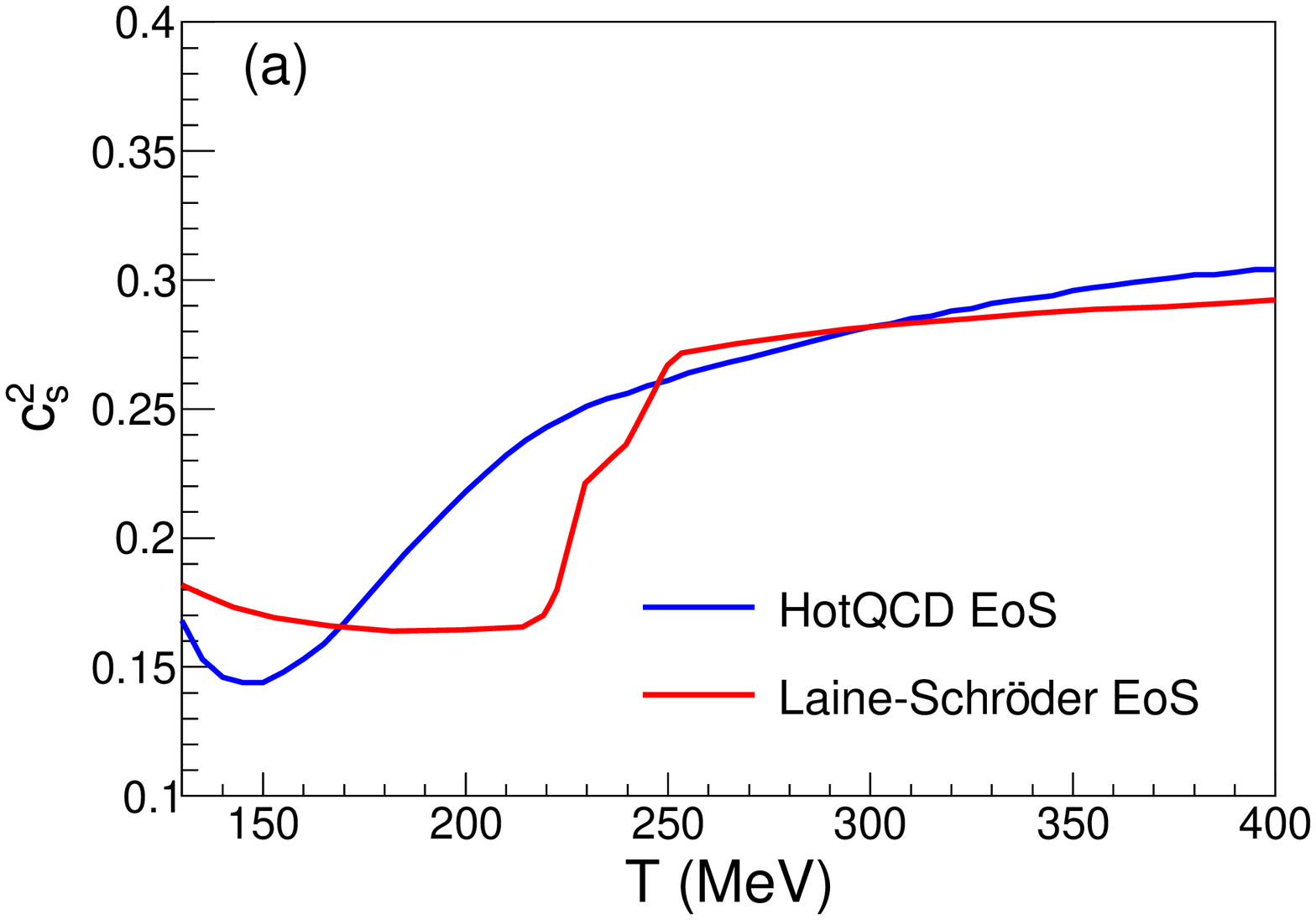}
\includegraphics[bb=0 0 567 409,width=0.49\textwidth]{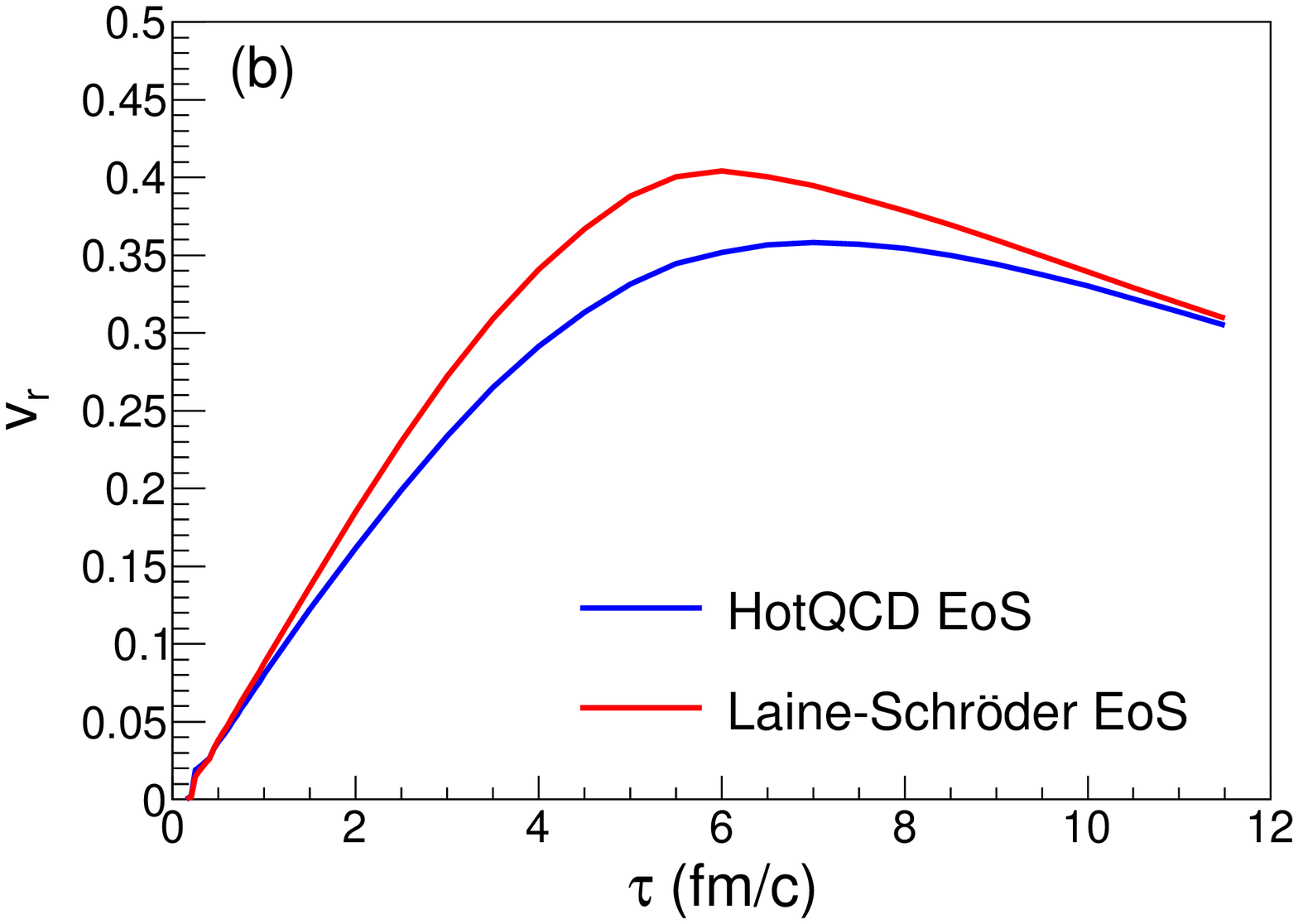}
\caption{The speed of sound square dependency on the temperature $c_s^2(T)$ (a) 
and the radial flow dependency on the proper time $v_r(\tau)$ at the
transverse radial coordinate $r_T=3$~fm (b), calculated within the iHKM  
for the two quark-gluon phase equations of state: the Laine-Schroeder 
EoS (red)~\cite{EoS} and the HotQCD Collaboration EoS (blue)~\cite{EoS2}.
\label{cst2}}
\end{figure}

In Fig.~\ref{eos2} one can see the graphical comparison of the two EoS, utilized
in our analysis. The two $p(\epsilon)$ plots look close, however  
a steeper growth of the energy density at high temperatures is observed
for the Laine-Schroeder EoS.
To analyze how the utilized EoS affects the magnitudes of the collective flows
at the hadronization hypersurface, in Fig.~\ref{cst2} we compare the dependencies
of the speed of sound square $c_s^2$ on the temperature $T$ and the radial 
flow~$v_r$ at $r_T=3$~fm on the proper time $\tau$ in iHKM for the two 
applied EoS's. From the first plot it follows, that at the temperatures,
close to $T_\mathrm{p}$, the $c_s^2$ magnitudes for the two equations of state
are different, which should result in different accelerations in hydrodynamical
regime. However, the second plot demonstrates, that the radial flow values
at the times near $\tau \approx 10$~fm/$c$, corresponding to the final stage of
the matter particlization, appear to be fairly close for both EoS's.

\begin{figure}
\centering
\includegraphics[bb=0 0 567 409, width=0.49\textwidth]{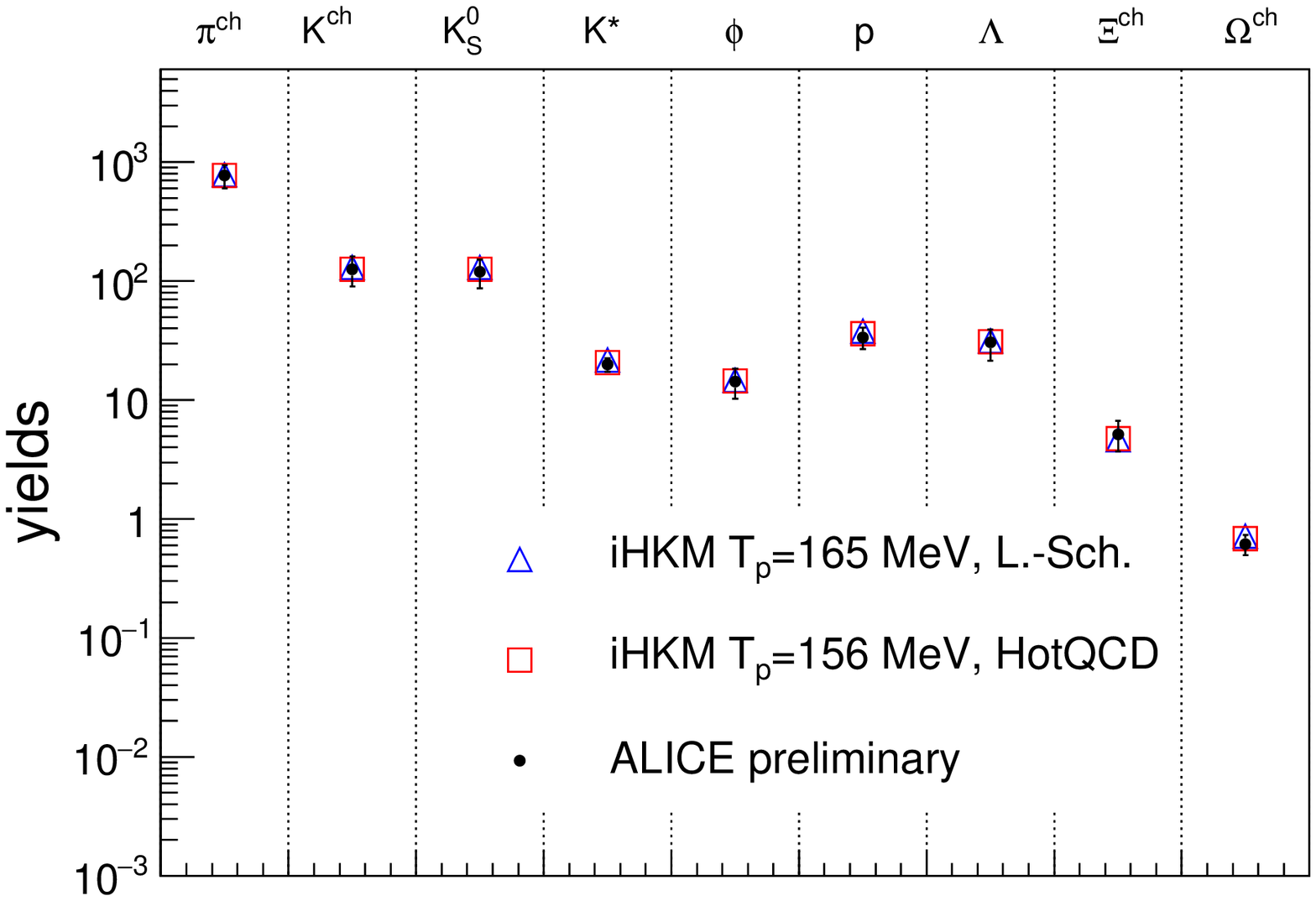}
\includegraphics[bb=0 0 567 409, width=0.49\textwidth]{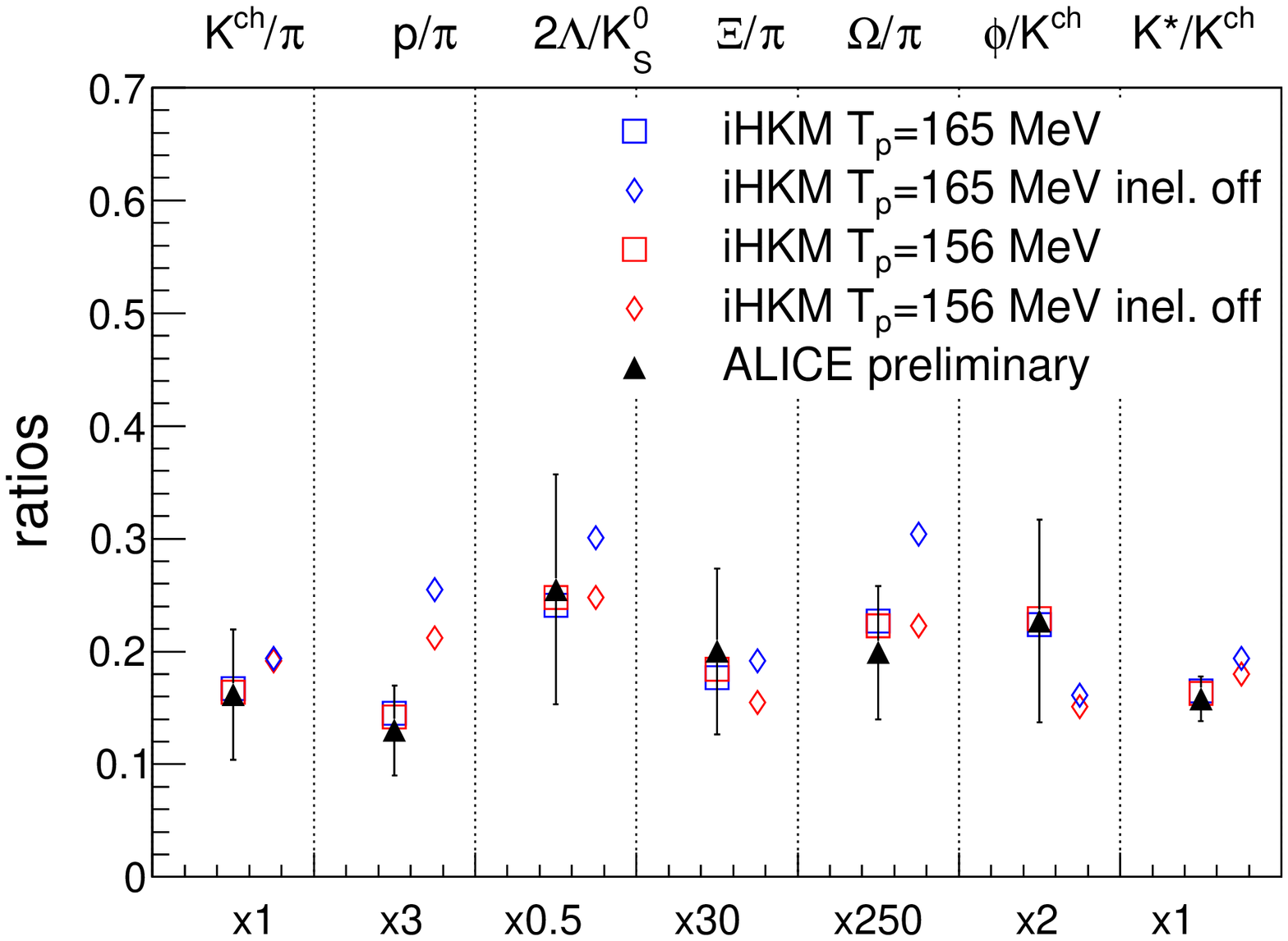}
\caption{The particle yields $dN/dy$ (left) and particle number ratios (right)
calculated in the iHKM using the two equations of state for quark-gluon matter
with the corresponding particlization temperatures: the Laine-Schroeder EoS
with $T_\mathrm{p}=165$~MeV and the HotQCD Collaboration EoS with $T_\mathrm{p}=156$~MeV.
The results concern Pb+Pb collisions at the LHC energy $\sqrt{s_{NN}}=5.02$~TeV, $c=0-10\%$
and are presented along with the ALICE Collaboration preliminary data~\cite{aliceqm1,aliceqm2}.
The particle number ratios were calculated in the two modes: the full iHKM simulation
and the simulation without inelastic reactions at the afterburner stage.
\label{yields}}
\end{figure}

\begin{figure}
\centering
\includegraphics[bb=0 0 567 409, width=0.49\textwidth]{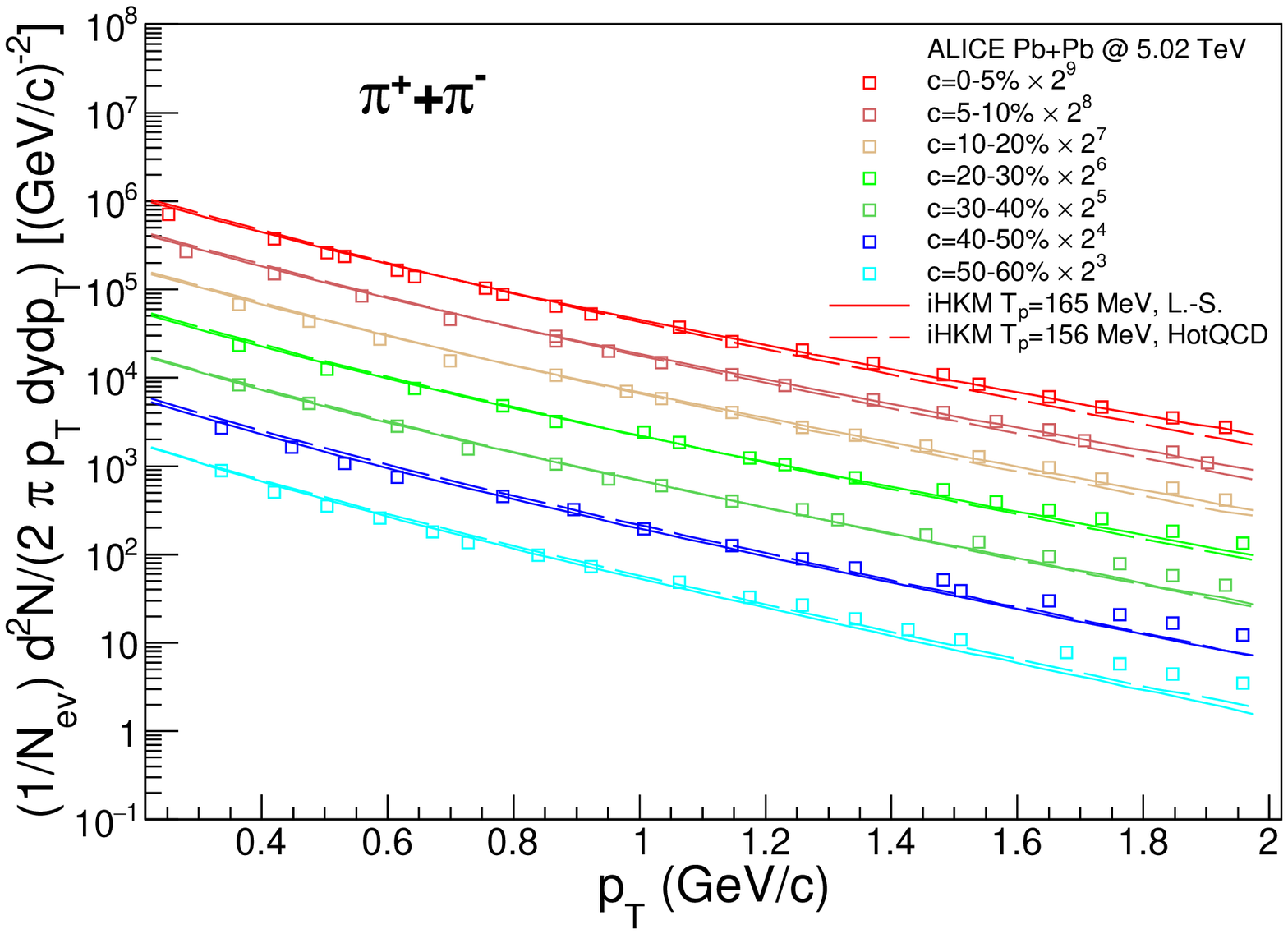} 
\includegraphics[bb=0 0 567 409, width=0.49\textwidth]{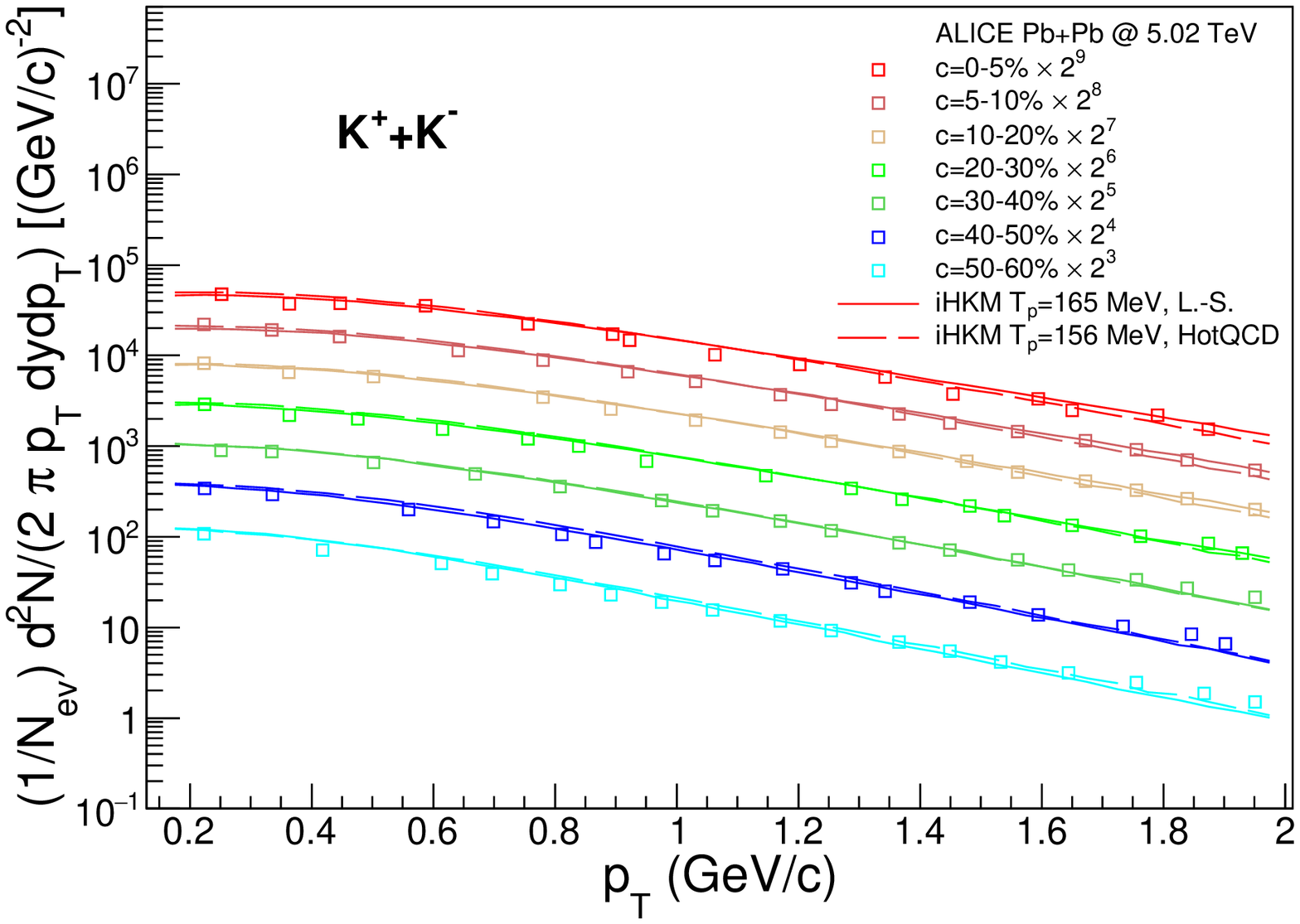} 
\caption{The $p_T$ spectra of pions (left) and kaons (right) for different centralities, 
calculated in the iHKM using the two equations of state for quark-gluon phase
with the corresponding particlization temperatures: the Laine-Schroeder EoS
with $T_\mathrm{p}=165$~MeV and the HotQCD Collaboration EoS with $T_\mathrm{p}=156$~MeV.
The model results are compared to the ALICE Collaboration preliminary data~\cite{spec502}. 
The data for different centrality classes are scaled by the different powers of 2
for better readability.
\label{specs}} 
\end{figure}

\begin{figure}
\centering
\includegraphics[bb=0 0 567 384, width=0.50\textwidth]{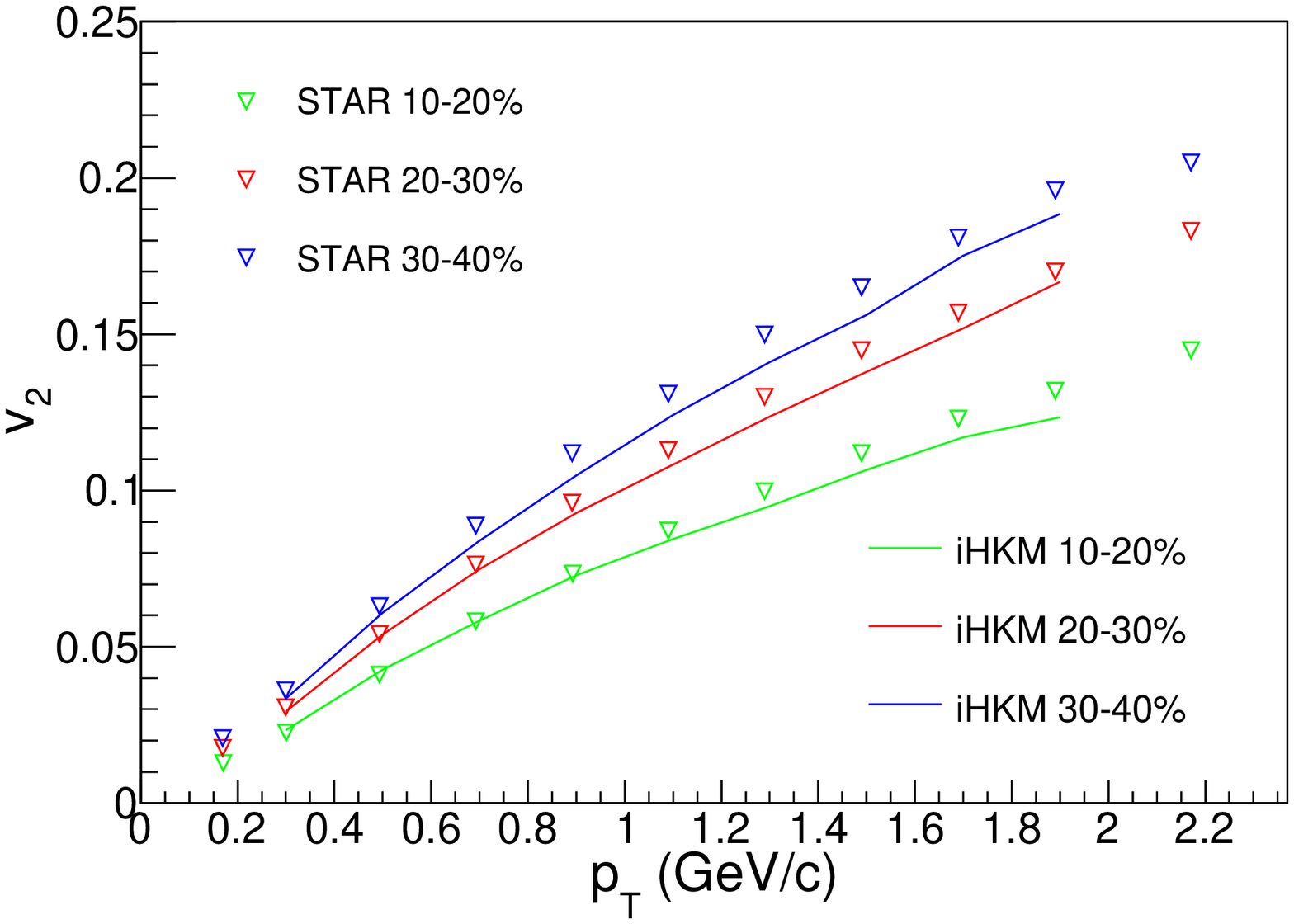}
\includegraphics[bb=0 0 567 409, width=0.48\textwidth]{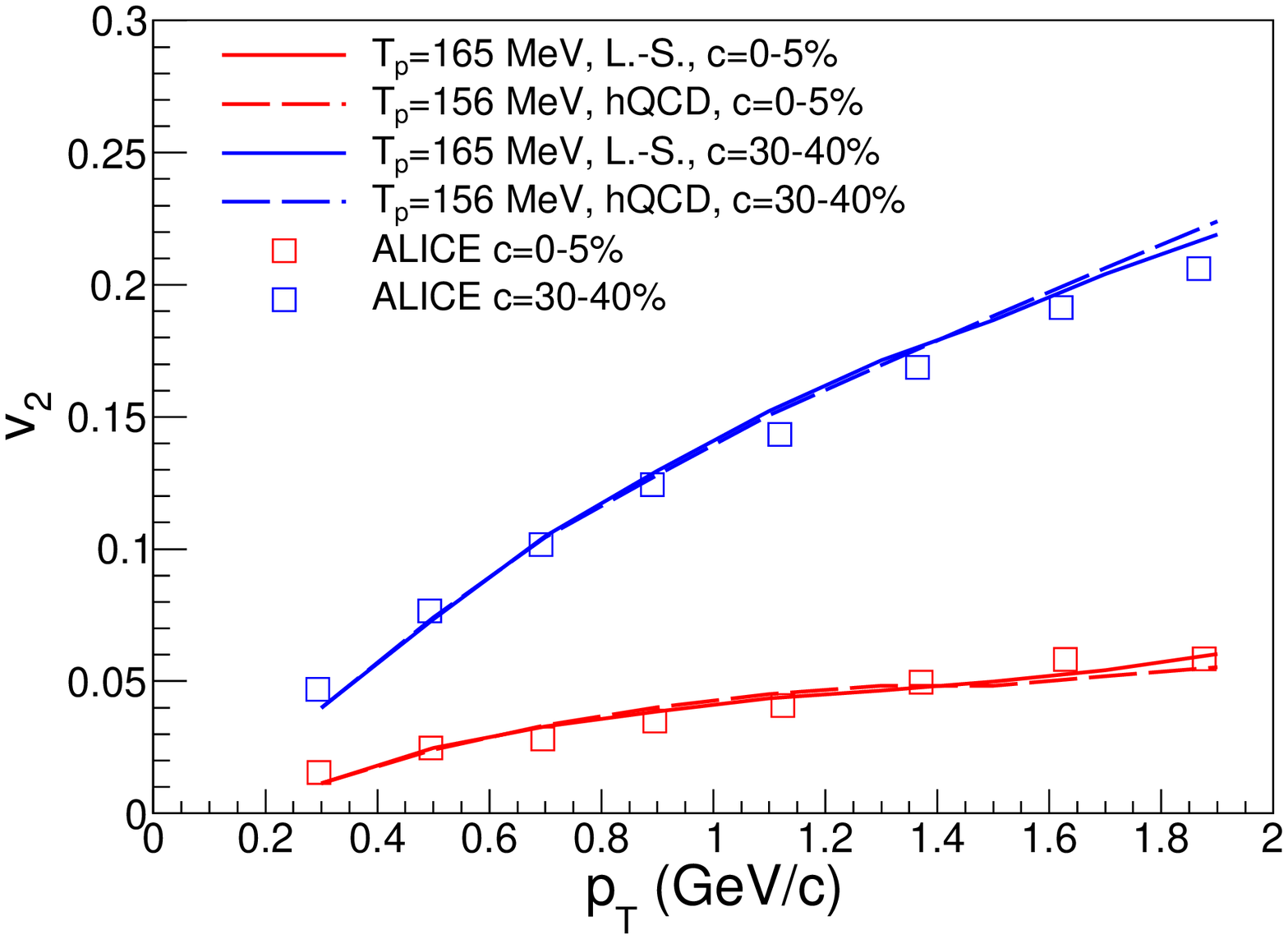}
\caption{The all charged particles $v_2(p_T)$ dependency, 
calculated in iHKM for the top RHIC energy Au+Au collisions (left) and for the LHC energy $5.02 A$~TeV 
Pb+Pb collisions (right). The different centrality classes are concerned ($c=10-20\%$, $c=20-30\%$ and $c=30-40\%$
in the RHIC case; $c=0-5\%$ and $c=30-40\%$ in the LHC case).
In the LHC case the two equations of state for quark gluon phase with their corresponding particlization 
temperatures were used for the simulations: the Laine-Schroeder EoS
with $T_\mathrm{p}=165$~MeV and the HotQCD Collaboration EoS with $T_\mathrm{p}=156$~MeV.
The experimental points correspond to the STAR and the ALICE Collaborations data, taken from \cite{v2star}
and \cite{v2alice} respectively.  
\label{v2}}
\end{figure}

\begin{figure}
\centering
\includegraphics[bb=0 0 567 409, width=0.49\textwidth]{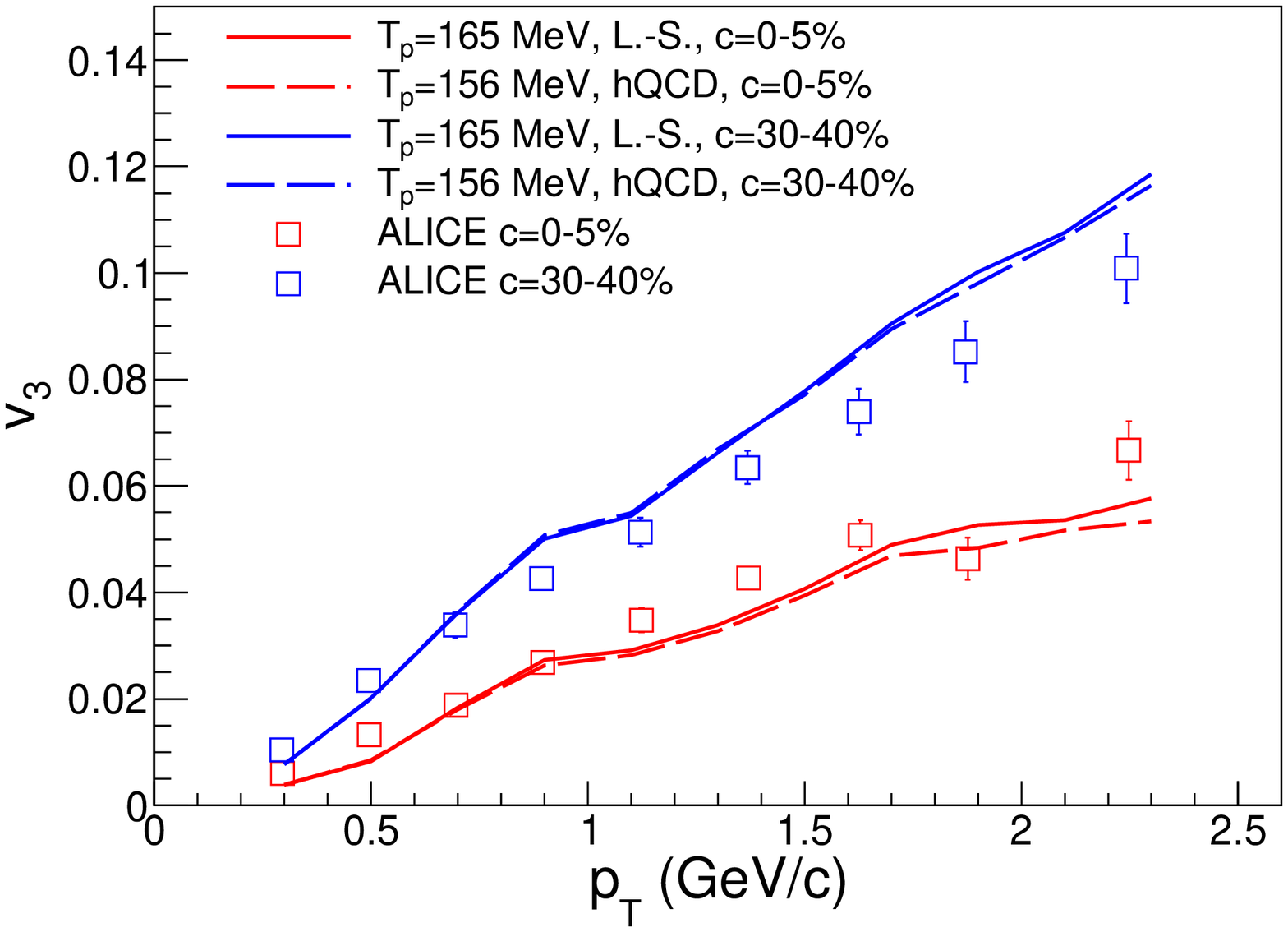}
\includegraphics[bb=0 0 567 409, width=0.49\textwidth]{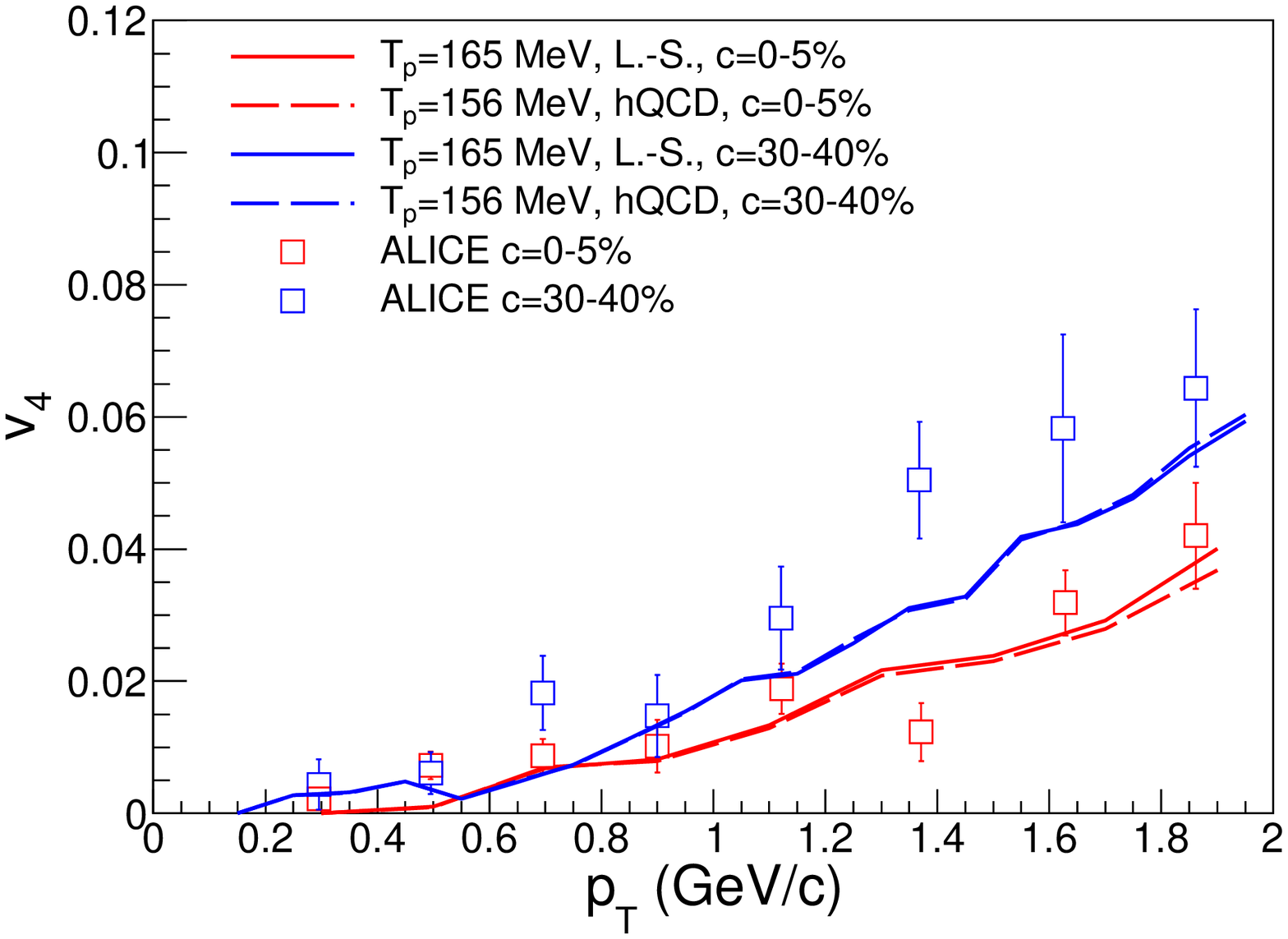} 
\caption{The all charged particles $v_3$ and $v_4$ flow harmonics for different $p_T$, calculated in iHKM
for Pb+Pb collisions at the LHC energy $5.02 A$~TeV.
The Laine-Schroeder EoS with $T_\mathrm{p}=165$~MeV and the HotQCD Collaboration EoS with $T_\mathrm{p}=156$~MeV
are used in the simulations for quark-gluon matter. The results for the two centrality classes, 
$c=0-5\%$ and $c=30-40\%$, are presented. The ALICE Collaboration experimental points are taken from \cite{v2alice}.  
\label{v3v4}} 
\end{figure}

\begin{figure}
\centering
\includegraphics[bb=0 0 567 409, width=0.88\textwidth]{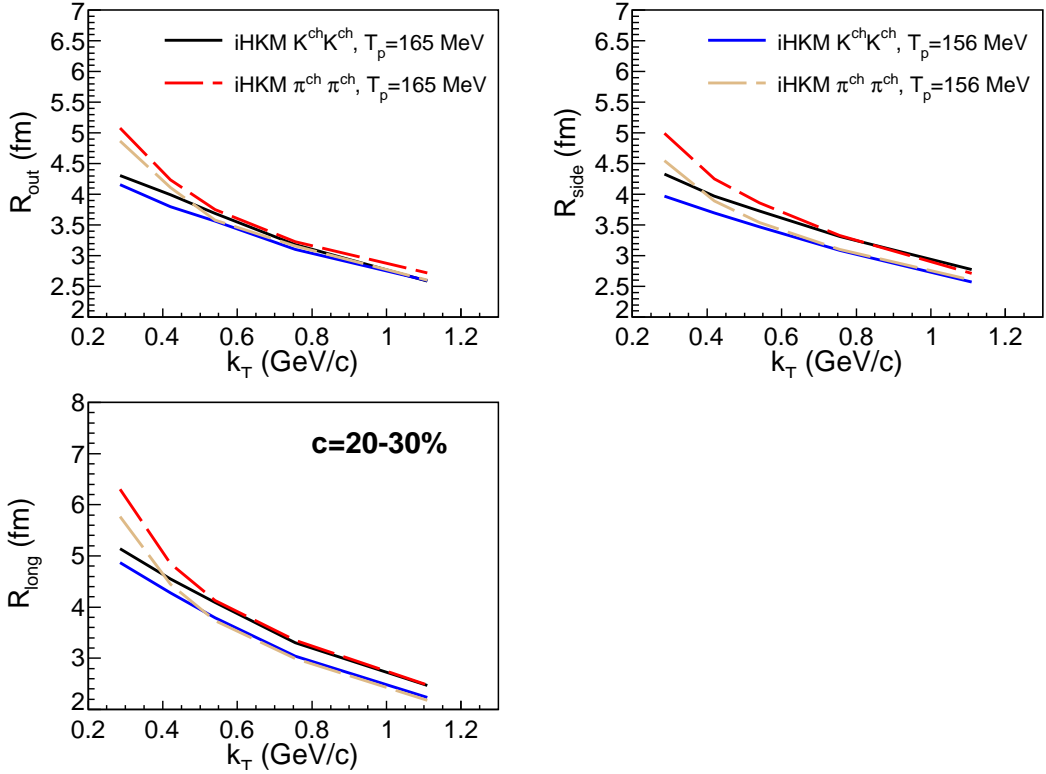}
\caption{The pion and kaon interferometry radii dependencies on $k_T$,
as predicted within iHKM for Pb+Pb collisions at the LHC energy $\sqrt{s_{NN}}=5.02$~TeV, $c=20-30\%$. 
The Laine-Schroeder EoS with $T_\mathrm{p}=165$~MeV and the HotQCD Collaboration EoS with $T_\mathrm{p}=156$~MeV
are used in the simulations for quark-gluon matter.
\label{rad2030}}
\end{figure}

The analysis, carried out in~\cite{ratiosour,lhc502-ihkm} shows that the experimental
particle yields and their ratios, transverse momentum spectra of identified
particles ($\pi$, $K$, $p$), $v_n$ flow harmonics (for $n=2,3,4$) and the femtoscopy
radii of pions and kaons in Pb+Pb collisions at the LHC energies can be simultaneously 
described in iHKM with an equally good accuracy, using both Laine-Schroeder and HotQCD 
Collaboration equations of state, if switching to another EoS is accompanied by re-tuning of
$\epsilon_0(\tau_0)$ parameters (see Figs.~\ref{yields}, \ref{specs}, \ref{v2},
\ref{v3v4}, \ref{rad2030}). The obtained results speak in favor of the hypothesis about
the continuous character of chemical and thermal freeze-outs in the process of heavy ion collisions.
Within the model it means, that the possible modifications in the system's expansion
as a continuous medium, caused by the change of the EoS and leading to the changes in the final
simulation results, can be compensated during the afterburner dynamics. This assumption
is supported by our results for particle number ratios, calculated in the two regimes:
the full iHKM simulation and the one with inelastic processes turned off at the afterburner stage (Fig.~\ref{yields}).
The baryonic ratios, calculated without inelastic processes, have different values for the different
utilized EoS's, however the same ratios, calculated in the full mode, coincide and agree with the experimental
data.

Indeed, the post-hydrodynamic stage of the matter evolution at the LHC energy scales has a long duration,
so that elastic and inelastic hadron-hadron interactions taking place at this stage should strongly
affect the formation of bulk observables, measured in the experiment and calculated in the simulations. 
In Refs.~\cite{alice1, alice2} the ALICE Collaboration pointed out, that it was the accounting for
the annihilation processes at the final stage of collision, that allowed to achieve a better agreement
of the model (anti)proton $p_T$ spectra and yields with the data in the HKM model~\cite{hkm3}. 
The influence of the afterburner inelastic interactions on the measured particle yields is 
investigated, e.g. in~\cite{beccatini}. 

From the theoretical point of view,
an abrupt transition from a very fast expansion of chemically equilibrated matter with intensive
inelastic reactions passing within the system to chemically frozen matter evolution without any 
inelastic processes, supposed by a sudden chemical freeze-out concept, looks unrealistic. 
The same concerns a sudden kinetic freeze-out. It would mean a sharp switching from a quite large
hadron-hadron interaction cross-section to a vanishing one. But what would be the physical reason for this?
Even a phase transition between quark-gluon matter and hadron gas in the considered expanding 
system can hardly be sudden in time.

In the paper~\cite{kstar}, analyzing in iHKM the $K^{*}(892)$ observability in view of the interaction 
of its decay daughters with the hadronic medium at the afterburner stage, we found that in case of 
the LHC Pb+Pb collisions at $2.76 A$~TeV the effective duration of the kinetic freeze-out can reach
about 5~fm/$c$ after the particlization. Note also, that the term ``continuous freeze-out'' does not
mean only different freeze-out times for different hadron species, as supposed, e.g. in~\cite{chat}.
The emission picture, obtained within iHKM in~\cite{kstar}, suggests also the continuous character of particle 
radiation for each given hadron sort.

\subsection{Particle production}
Once the model is calibrated for description of a particular collision type, based on the
experimental charged particle multiplicity dependency on collision centrality and pion $p_T$ spectrum
slope in the most central events, a wide class of soft observables, including particle yields,
particle number ratios, transverse momentum spectra, flow harmonics, femtoscopy scales, can be
described simultaneously within iHKM, using the single found setting of model parameters for each 
experiment set-up (again see Figs.~\ref{yields}, \ref{specs}, \ref{v2}, \ref{v3v4}, \ref{rad2030}).
The results, obtained in~\cite{ihkm2,ratiosour,lhc502-ihkm,rhic-ihkm}, provide a detailed
description of particle production in Au+Au collisions at the RHIC energy $\sqrt{s_{NN}}=200$~GeV and 
in Pb+Pb collisions at the two LHC energies, $\sqrt{s_{NN}}=2.76$~TeV and $\sqrt{s_{NN}}=5.02$~TeV.

\begin{figure}
\centering
\includegraphics[bb=0 0 567 383, width=0.49\textwidth]{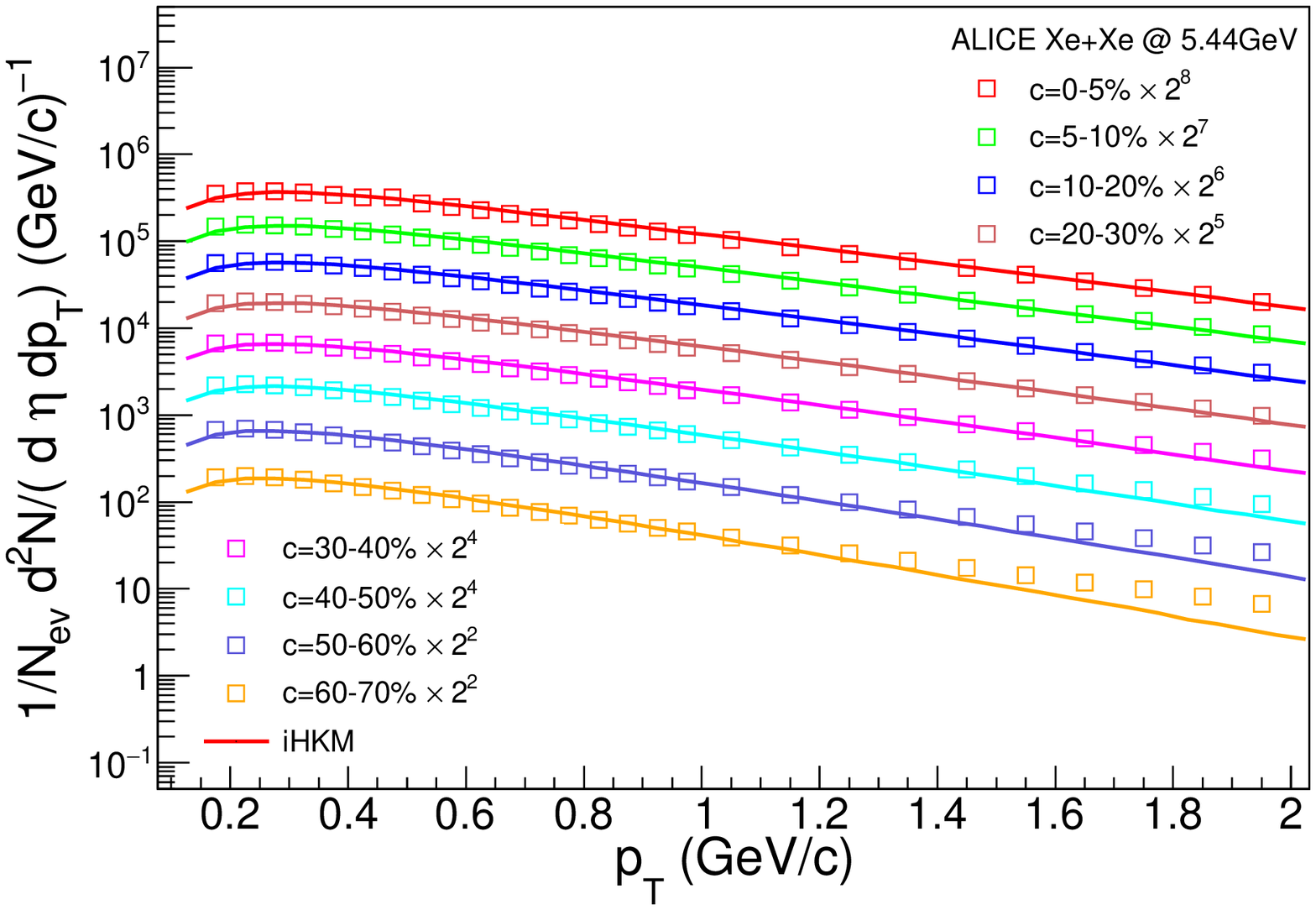}
\includegraphics[bb=0 0 567 383, width=0.49\textwidth]{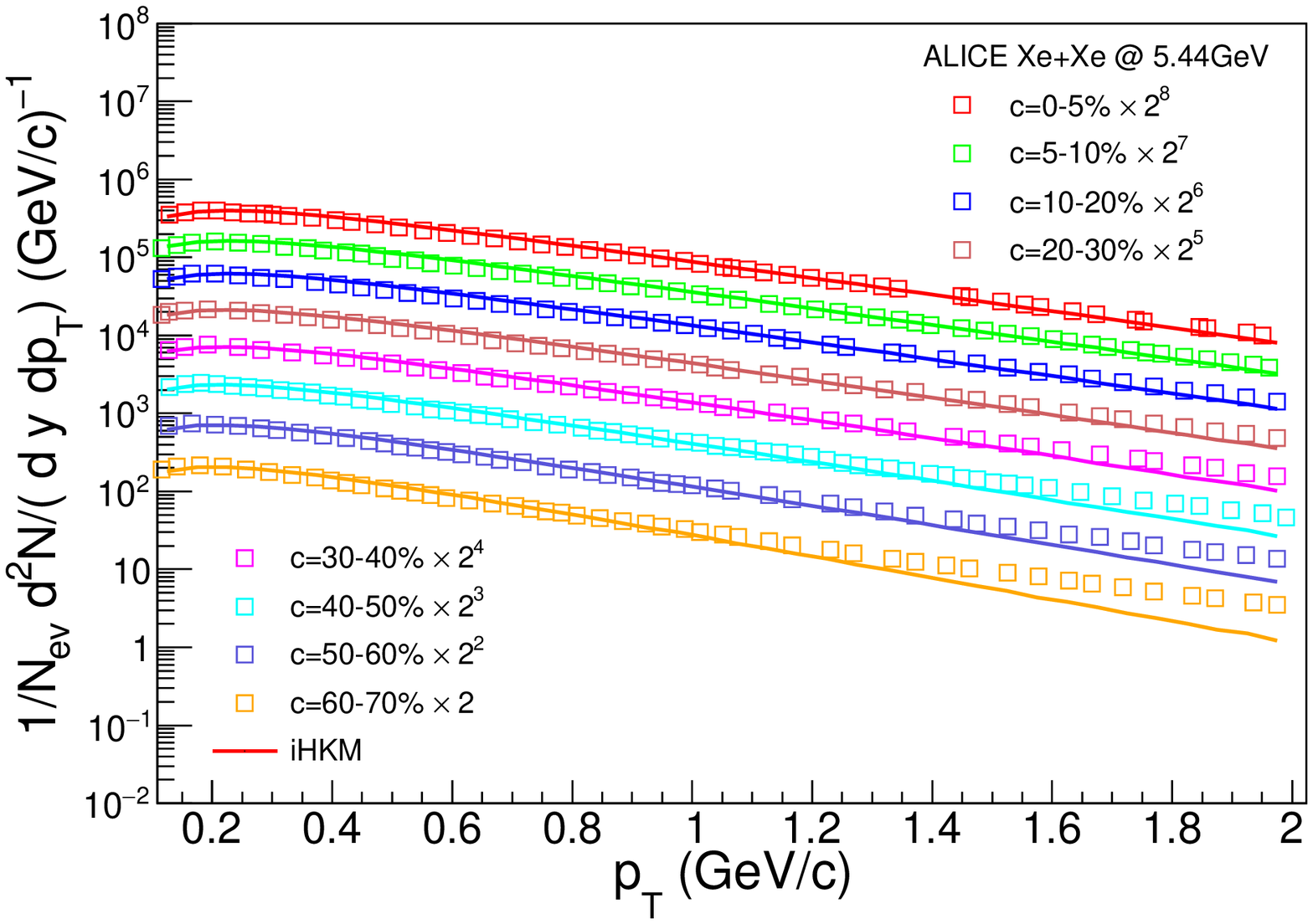}
\caption{The $p_T$ spectra of all charged particles ($|\eta|<0.8$, left) and pions ($|y|<0.5$, right) 
in the LHC Xe+Xe collisions at $5.44 A$~TeV, calculated in iHKM, 
in comparison with the ALICE Collaboration experimental points~\cite{alicexe,alicexe2}. 
\label{all2}} 
\end{figure}

\begin{figure}
\centering
\includegraphics[bb=0 0 567 383, width=0.49\textwidth]{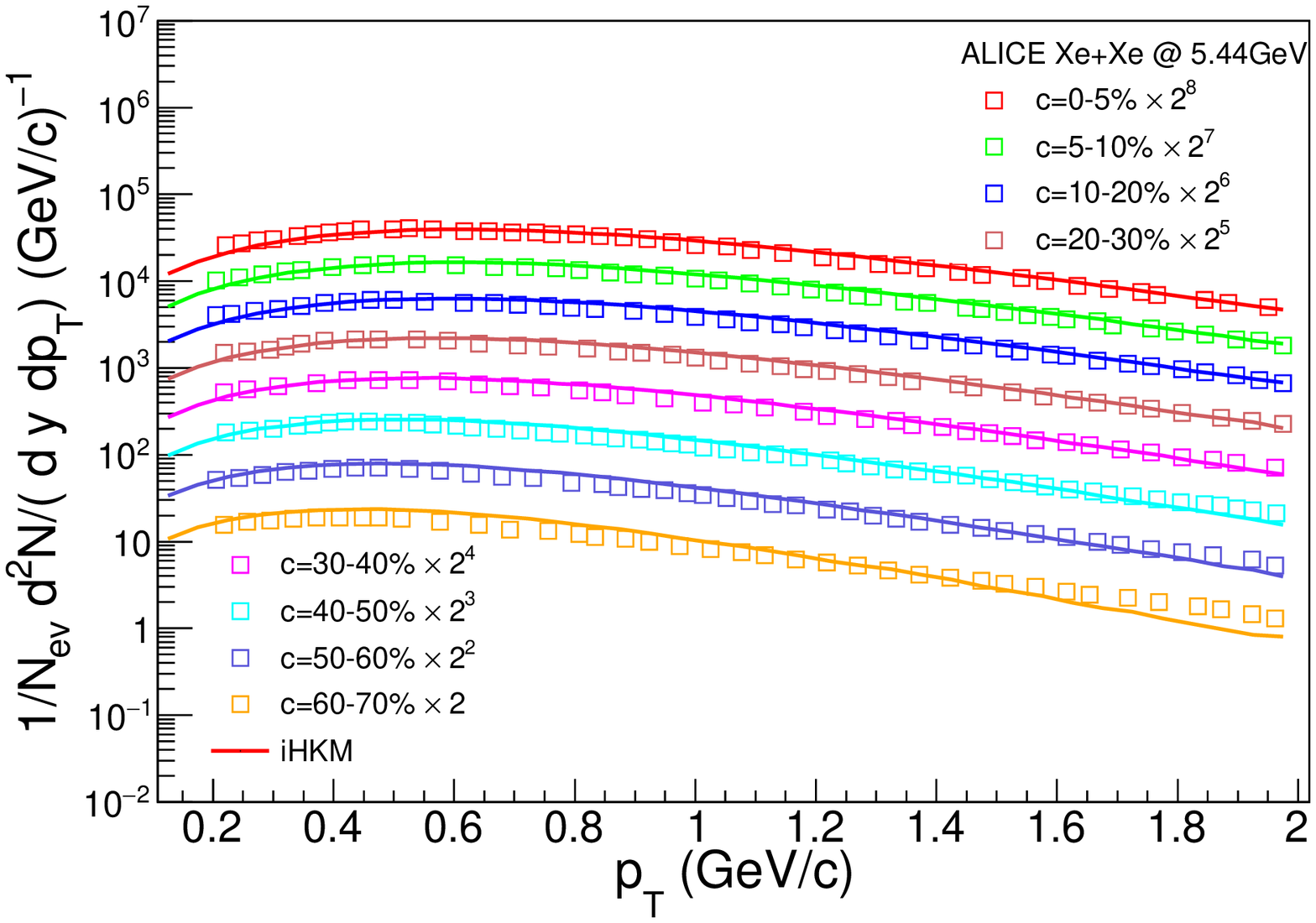}
\includegraphics[bb=0 0 567 383, width=0.49\textwidth]{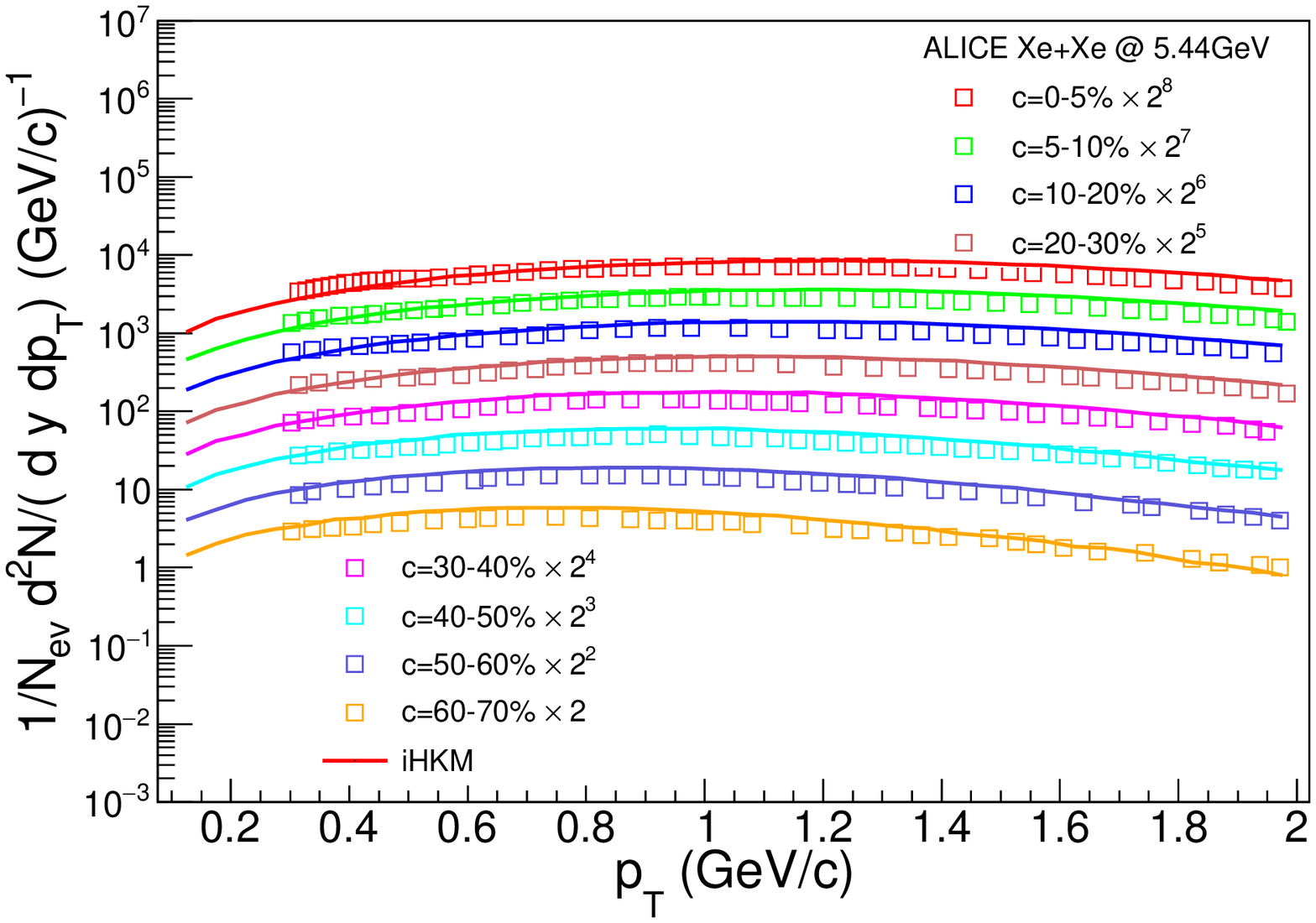}
\caption{The same as in Fig.~\ref{all2} for kaons (left) and protons (right), $|y|<0.5$ 
(The ALICE Collaboration experimental data are from the Ref.~\cite{alicexe2}).
\label{k2}} 
\end{figure}

In Figs.~\ref{all2}, \ref{k2} one can also see the latest iHKM results for $p_T$ spectra
of all charged particles, pions, kaons and protons in Xe+Xe collisions at the LHC energy $\sqrt{s_{NN}}=5.44$~TeV
for eight centrality classes ($c=0-5\%$, $c=5-10\%$, $c=10-20\%$, $c=20-30\%$, $c=30-40\%$, $c=40-50\%$, $c=50-60\%$ $c=60-70\%$).
The simulation results are compared with the ALICE Collaboration experimental data~\cite{alicexe,alicexe2}
and appear to reproduce them well in a wide $p_T$ range, that includes soft momentum interval.
The accuracy of the spectra description is comparable with previous iHKM results for the top RHIC energy
Au+Au collisions and the LHC Pb+Pb collisions.

\begin{figure}
\centering
\includegraphics[bb=0 0 567 409, width=0.86\textwidth]{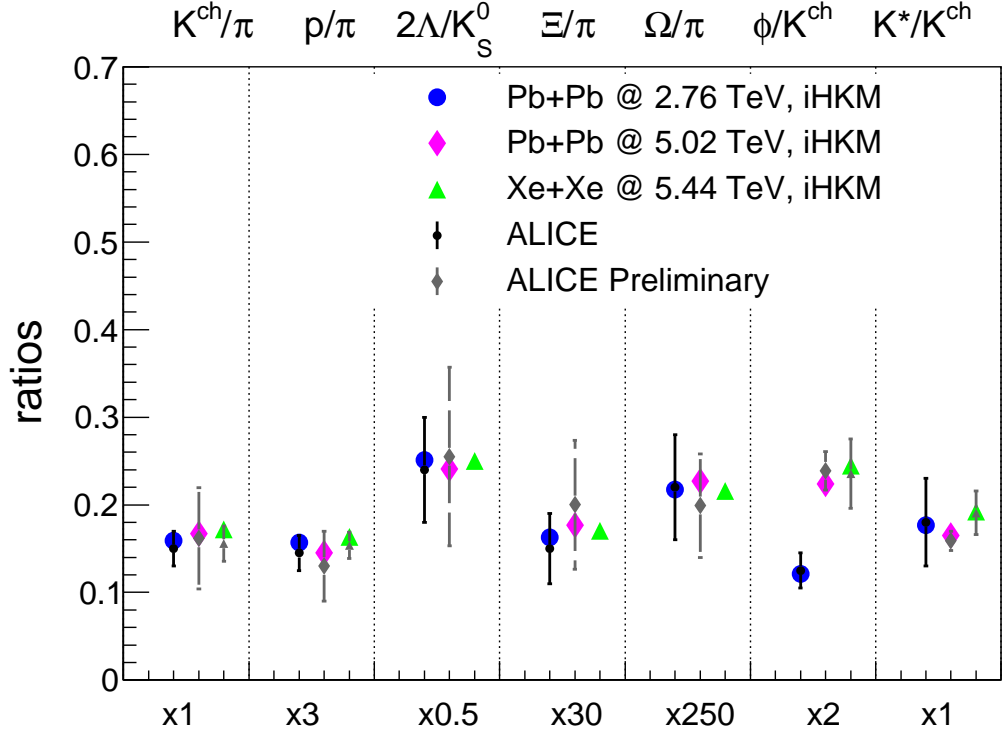}
\caption{The particle number ratios, calculated in iHKM for the LHC
Pb+Pb collisions at $2.76 A$~TeV and $5.02 A$~TeV, as well as for the LHC Xe+Xe collisions at  
$5.44 A$~TeV. The experimental data are also presented in cases when they are available~\cite{RatiosExp,aliceqm1,aliceqm2}.
\label{pnr2}} 
\end{figure}

The iHKM results for various particle number ratios in the LHC Xe+Xe collisions at $5.44 A$~TeV are presented 
in Fig.~\ref{pnr2} together with the results for other LHC collisions and the corresponding experimental 
points~\cite{RatiosExp,aliceqm1,aliceqm2}. The shown model ratios are obtained in the full simulation regime,
with inelastic processes at the post-hydrodynamic stage turned on. They agree with the measured data within the
errors. The ratios of the same hadron yields in different collisions are close to each other, except for
the $\phi/K^\mathrm{ch}$ ratio in Pb+Pb collisions at the LHC energy $\sqrt{s_{NN}}=2.76$~TeV.
The latter fact can be possibly connected with a shorter in time and not so intensive afterburner dynamics
in the $2.76 A$~TeV case as compared to higher energy collisions at $\sqrt{s_{NN}}=5.02$~TeV and 
$\sqrt{s_{NN}}=5.44$~TeV.
The afterburner dynamics can be important for the $\phi/K^\mathrm{ch}$ ratio formation, since additional 
$\phi$ resonances appear at the late stage of the collision due to $K^{+}K^{-}$ recombinations and 
correlations~\cite{kstar}, increasing both the $\phi$ yield and the $\phi/K$ ratio. This effect should be 
smaller for collisions at smaller energy, so that the $\phi/K$ ratio in $2.76 A$~TeV collisions will be 
smaller than those in $5.02 A$~TeV and $5.44 A$~TeV cases.

In order to describe the particle production in Au+Au collisions at the top RHIC 
energy~\cite{rhic-ihkm} we had to introduce a downscaling strangeness suppression
factor $\gamma_s$ for non-central events. It was assumed to depend on the particlization
time $\tau_\mathrm{p}$, calculated in iHKM and approximately characterizing the life-time of quark-gluon matter 
for each collision centrality. 
\begin{equation} 
\gamma_s(\tau_\mathrm{p}) = A \exp(-b/\tau_\mathrm{p}),
\label{gammas} 
\end{equation}
where $A=1.1$ and $b=0.8$~fm/$c$.

The utilization of such a factor is motivated by the consideration that in non-central events
the strange quarks do not have enough time to reach the local chemical equilibrium.
This leads to a noticeable lowering of kaon and proton spectra (about a half of observed protons 
are produced as a result of strange resonances decays). The hyperon spectra are also affected.
To correct the situation, each calculated hadron yield was multiplied by $\gamma_s(\tau_\mathrm{p})^{S_i}$, 
where $S_i$ is the $i^\mathrm{th}$ hadron sort strangeness. As a result, a good description for the measured
kaon, proton and hyperon $p_T$ spectra, as well as for the corresponding particle number ratios was reached.

\begin{figure}
\includegraphics[width=0.49\textwidth]{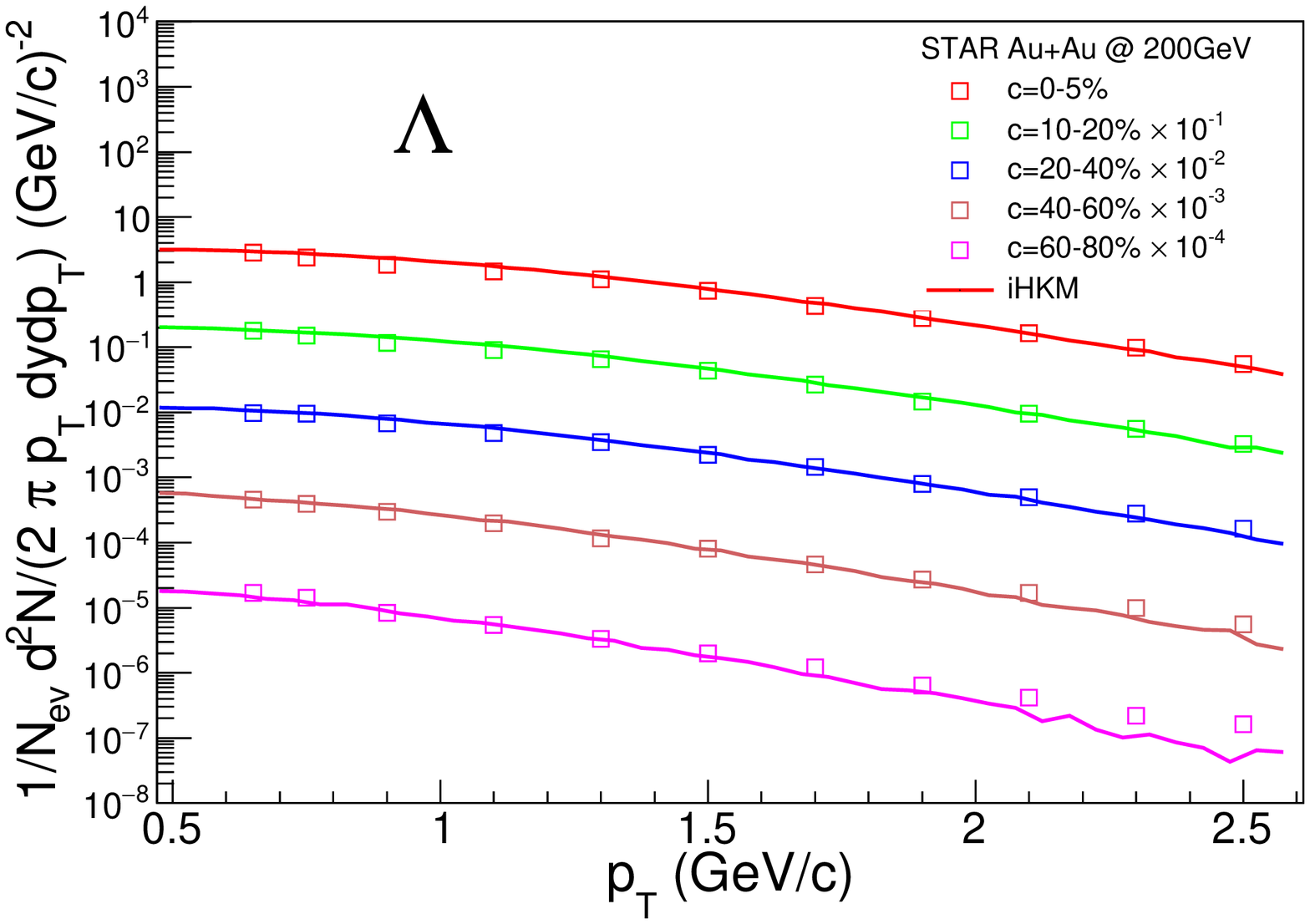}
\includegraphics[width=0.49\textwidth]{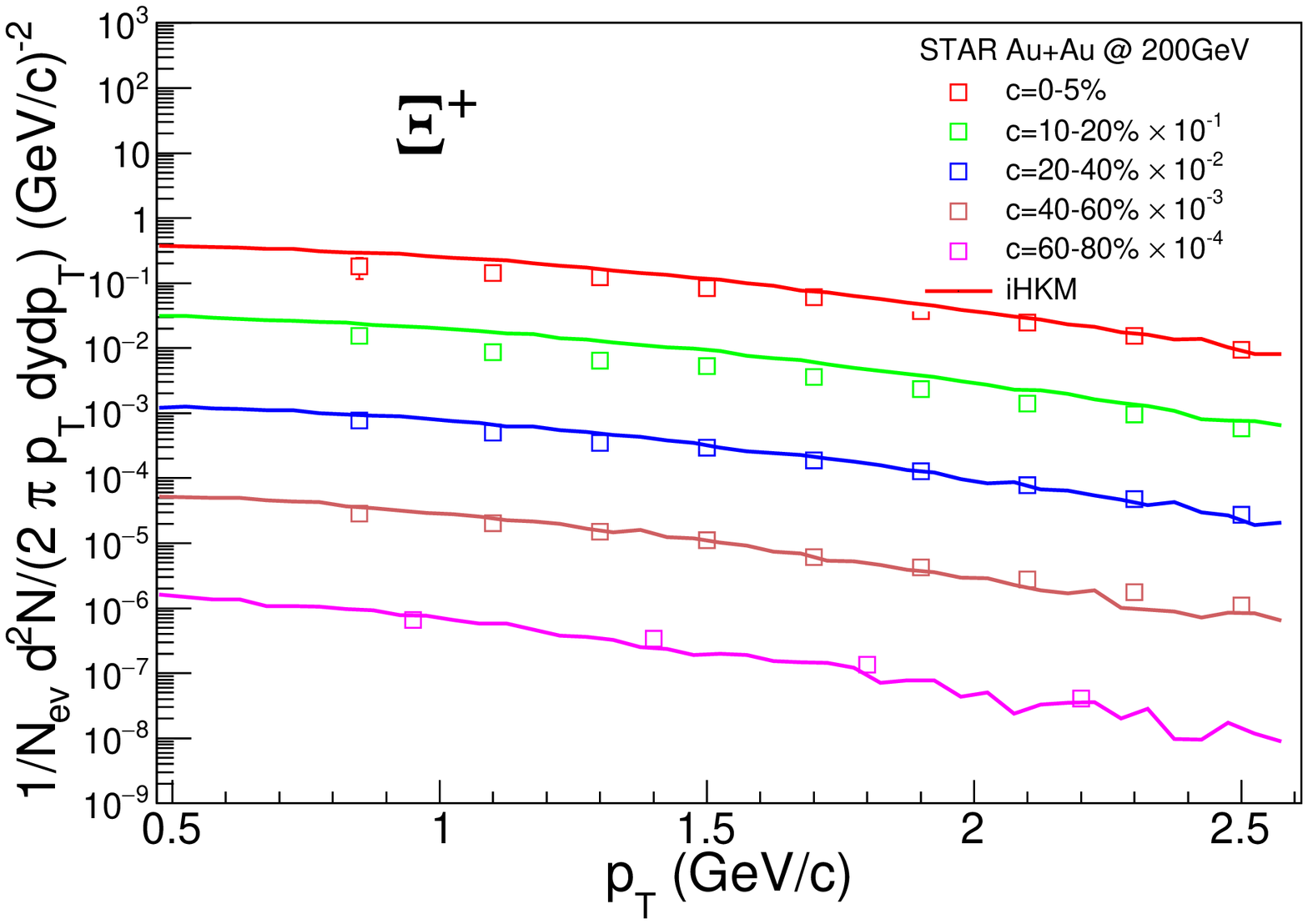}
\caption{The iHKM results for $\Lambda$ and $\Xi^{+}$ hyperons $p_T$ spectra in comparison with 
the STAR Collaboration experimental data~\cite{star_bar}. The lambdas with $|y|<1$ and cascade particles with
$|y|<0.75$ were selected for the analysis.}
\label{hyper1}
\end{figure}

\begin{figure}
\includegraphics[width=0.49\textwidth]{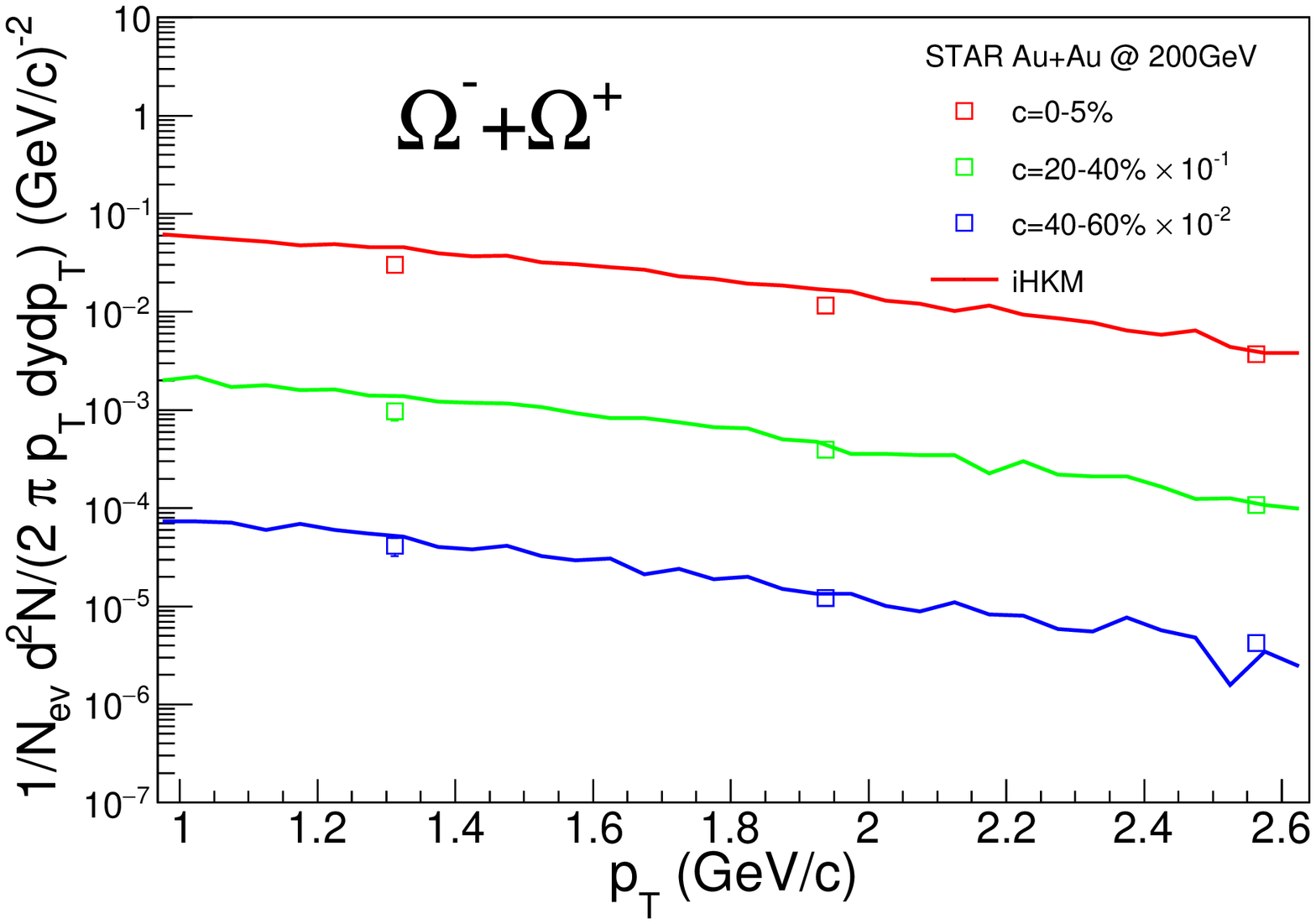}
\includegraphics[width=0.49\textwidth]{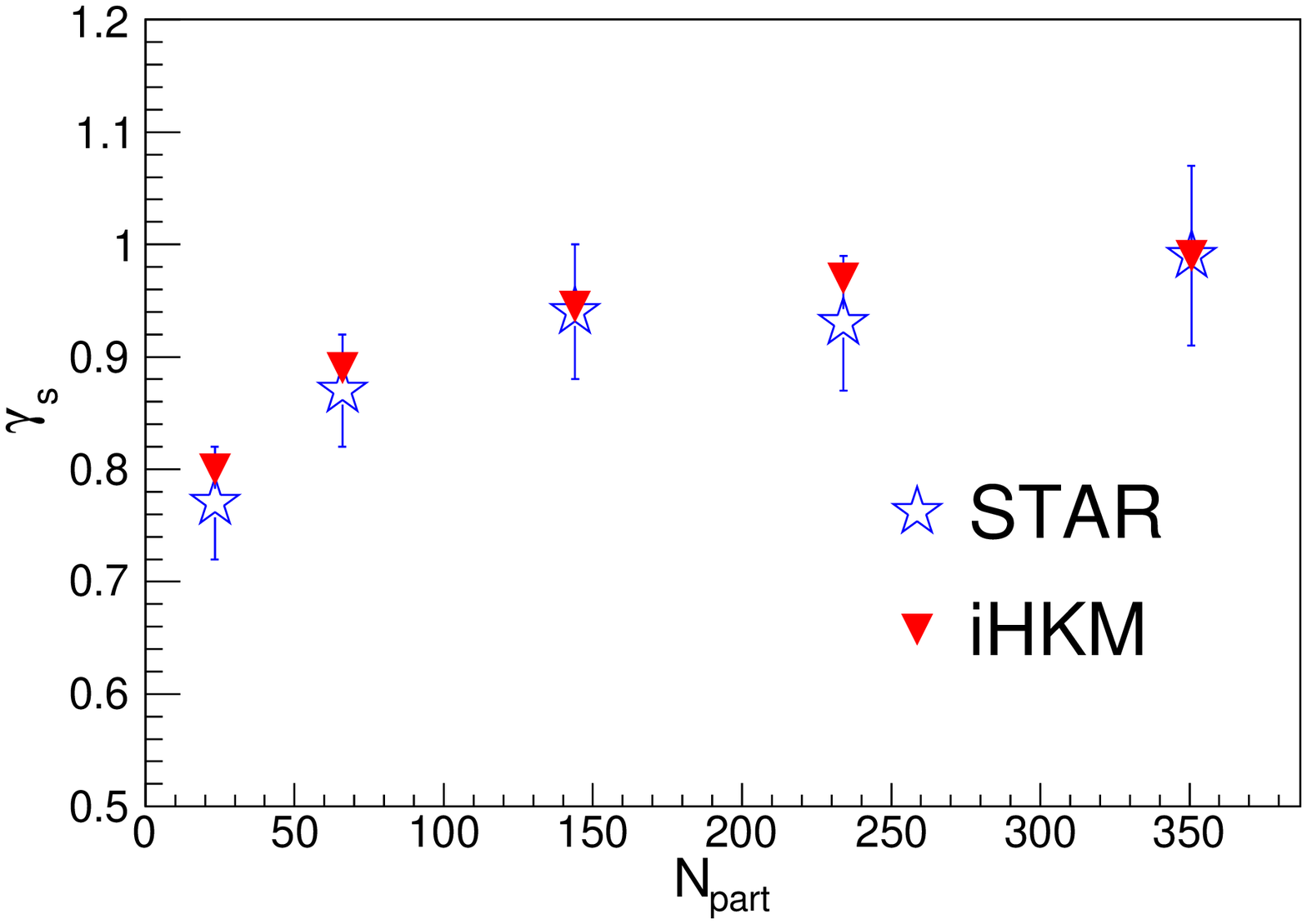}
\caption{The iHKM results for $\Omega$ hyperon $p_T$ spectra (for $|y|<0.75$) in comparison with 
the STAR Collaboration experimental data~\cite{star_bar} (left) and the $\gamma_s$ factors, applied
in iHKM simulations, together with those extracted by STAR in~\cite{star_bar}.}
\label{hyper2}
\end{figure}

In Figs.~\ref{hyper1}, \ref{hyper2} the results for $\Lambda$, $\Xi$ and $\Omega$ hyperon spectra are demonstrated,
as well as the comparison of $\gamma_s$ values, used in the model, and those extracted within the experimental
analysis from the Boltzmann fits to the measured spectra~\cite{star_bar}. The model curves for hyperon 
spectra agree with the data and the utilized $\gamma_s$ factors coincide with the experimental ones within the error
bars.


\subsection{Femtoscopy scales}
The correlation femtoscopy is a powerful experimental technique, that allows to reveal
the space-time structure of the system, formed in relativistic heavy-ion collision,
presenting it in terms of the interferometry (femtoscopy) radii.
The latter are extracted from the Gaussian fits $[1+\lambda\exp{(-q_i^2R_i^2)}]$ to the experimentally measured
two-particle momentum correlation functions $C(k_T,q)$ of identical particles in each $k_T$ bin.
Each $R_i(k_T)$ is commonly interpreted as the homogeneity length of the system in $i^\mathrm{th}$ direction, 
related to the region, where the particles with momenta close to $k_T$ are mostly emitted 
from~\cite{Sin1,Sin2,AkkSin2}.

The iHKM provides a satisfactory description of the femtoscopy radii in the considered high-energy collision
experiments, for which the experimental data are available for comparison (except, maybe, for a somewhat 
overestimated pion $out$ radii in the top RHIC energy Au+Au collisions at $c=0-10\%$~\cite{rhic-ihkm}, as compared
to the PHENIX Collaboration results), and allows to obtain predictions for not yet measured data 
(see Fig.~\ref{rad2030}). 

The observed in iHKM behaviour of the radii with $k_T$ is quite similar at all the collision energies.
At the fixed centrality the radii values grow by $15-20\%$, when going from the RHIC to the LHC, 
and the radii, related to the different LHC energies are very close, with differences not exceeding $3-5\%$.
Such similarity of the femtoscopy scales for the LHC collisions, that takes place despite the differences
in the system geometric sizes at the freeze-out stage, can be explained by the fact, that collective flow
velocity gradients are higher at larger energies, and so this compensates the different geometry influence,
resulting in close lengths of homogeneity in the considered cases~\cite{Sin1,AkkSin2}. 

An approximate $k_T$ scaling, that can be observed in Fig.~\ref{rad2030} 
between kaon and pion femtoscopy radii at $k_T>0.4$~GeV/$c$,
was firstly noticed for the LHC Pb+Pb collisions at $\sqrt{s_{NN}}=2.76$~TeV in iHKM 
simulations~\cite{ourkt} and then confirmed in the experimental analysis~\cite{alice-scaling}.

\begin{figure}
\centering
\includegraphics[bb=0 0 567 410, width=0.86\textwidth]{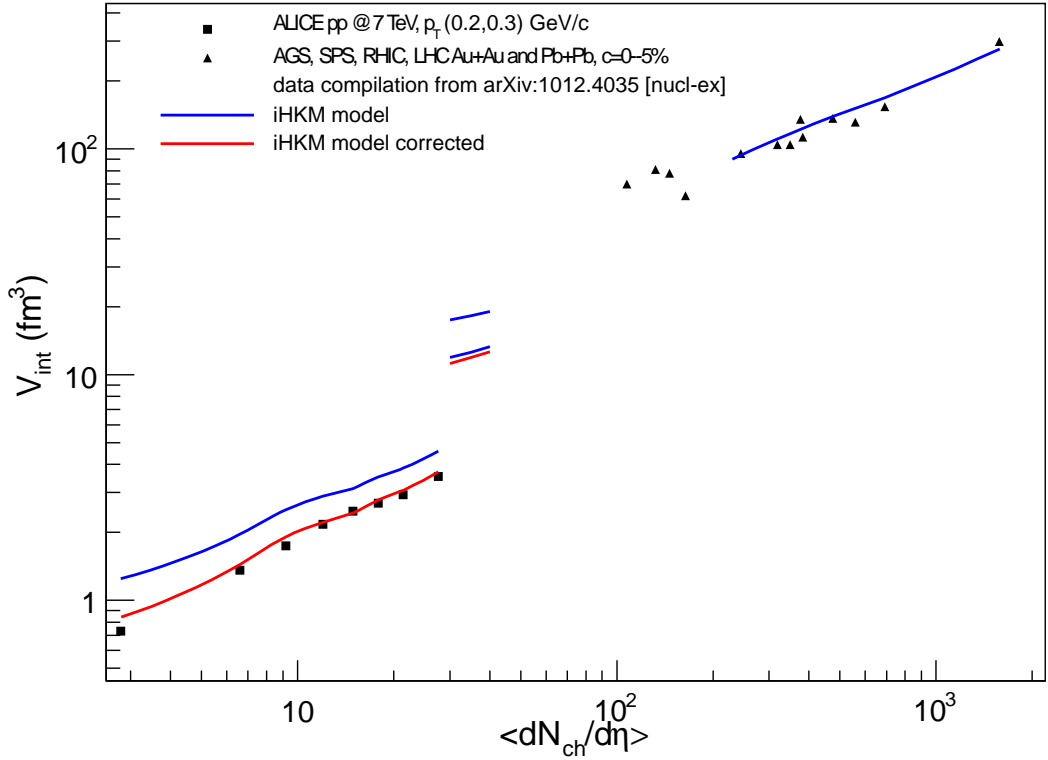}
\caption{The interferometry volume 
$V_{\mathrm{int}}=R_{\mathrm{out}}R_{\mathrm{side}}R_{\mathrm{long}}$ 
dependency on the mean charged particle density $\langle dN_{\mathrm{ch}}/d\eta \rangle$. 
The blue line at the right corresponds to iHKM results for the central heavy-ion collisions
at AGS, SPS, RHIC and LHC. The blue and red lines at the left describe the model results
for the LHC $p+p$ collisions at the energy $\sqrt{s}=7$~TeV.  
The corresponding experimental data are taken from the papers~\cite{alice_pp, AaExp1, AaExp2, 
AaExp3,AaExp4,AaExp5,AaExp6,AaExp7,AaExp8}.  
The line fragments in the middle correspond to the model predictions 
for the case of $p+$Pb collisions at the LHC energy $\sqrt{s}=5.02$~TeV,
calculated using the two different initial transverse system's sizes: 
$R=1.5$~fm (upper line) and $R=0.9$~fm (two lower lines).
The red lines show the results after application of the quantum corrections following~\cite{pp-lhc},
and the blue lines demonstrate the results without corrections.
\label{vint01}} 
\end{figure}

In the current paper we also present the comparison of the interferometry scales,
obtained in iHKM simulations for the high-energy collision events with close charged particle
multiplicities. Such a comparison has an aim to check the so-called ``scaling hypothesis''
(see, e.g.~\cite{Lisa1,Lisa2}), supposing that the interferometry radii and the corresponding
interferometry volume $V_{\mathrm{int}}=R_{\mathrm{out}}R_{\mathrm{side}}R_{\mathrm{long}}$ 
grow almost linearly with the mean charged particle density $\langle dN_{\mathrm{ch}}/d\eta \rangle$.
Despite the fact the scaling hypothesis is often assumed in femtoscopy studies,
we believe, that the femtoscopy scales should depend also on geometrical sizes
of the colliding ions (see Fig.~\ref{vint01}), as it was previously noticed in~\cite{scaling1,scaling2,pp-lhc}.

\begin{table}
\centering
\begin{tabular}{|c|c|c|}
 \hline
 Experiment & Centrality (\%) & $\langle dN_{\mathrm{ch}}/d\eta \rangle$ \\ 
 \hline
 Au+Au @ 200 GeV  & $0-5$   & 688	  \\
 Pb+Pb @ 2.76 TeV & $19-28$ & 693 	  \\
 Pb+Pb @ 5.02 TeV & $23-33$ & 677 	  \\
 Xe+Xe @ 5.44 TeV & $10-19$ & 680 	  \\
 \hline
\end{tabular}
\caption{The centrality and the mean charged particle multiplicity 
$\langle dN_{\mathrm{ch}}/d\eta \rangle$ at $|\eta|<0.5$ of the events,
chosen for the analysis of the femtoscopy scales dependency on the geometrical
sizes of the colliding nuclei.
\label{tab2}}
\end{table}

\begin{figure}
\centering
\includegraphics[bb=0 0 567 383, width=0.84\textwidth]{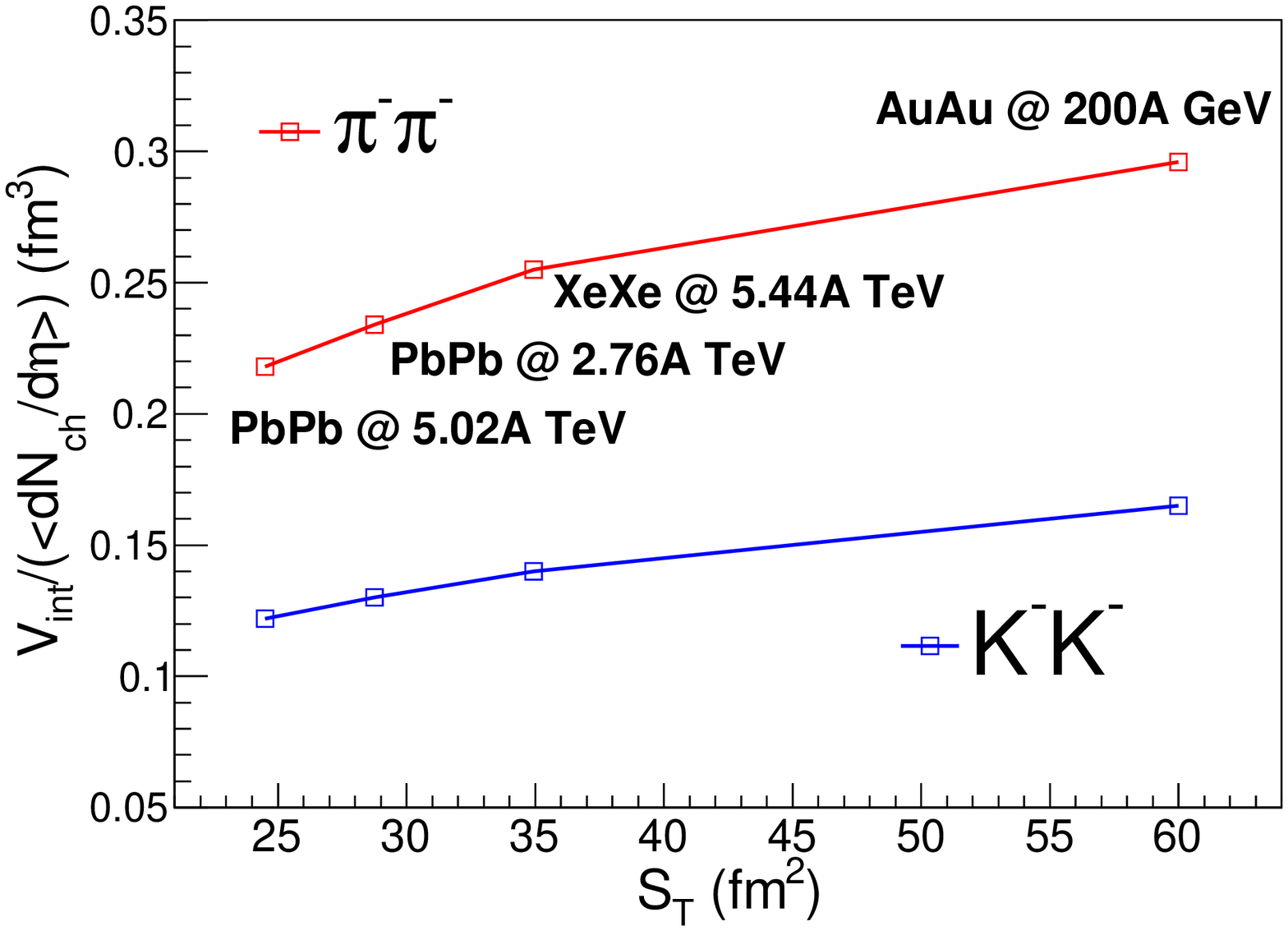}
\caption{The pion and kaon interferometry volume 
$V_{\mathrm{int}}=R_{\mathrm{out}}R_{\mathrm{side}}R_{\mathrm{long}}$,  
divided by the mean charged particle density $\langle dN_{\mathrm{ch}}/d\eta \rangle$, 
calculated in iHKM for different relativistic heavy-ion collisions, characterized 
by the areas $S_\mathrm{T}$ of the colliding nuclei initial transverse overlapping.
The corresponding collision centralities can be found in Table~\ref{tab2}.
The interferometry radii correspond to the pair transverse momentum $0.2<k_T<0.3$~GeV/$c$.
\label{vint1}} 
\end{figure}

To investigate this issue, for all the considered collision types we selected
the events of different centralities, such that the mean charged particle multiplicities
in each case were nearly equal (see Table~\ref{tab2}).
Then we built the graphs of pion and kaon interferometry volume dependencies on 
$S_\mathrm{T}$ parameter, which characterizes the sizes of the initial transverse
overlapping of the two colliding nuclei (see Fig.~\ref{vint1}). This parameter is calculated as
the area of nearly elliptical cross-section of the initial energy-density 
profile at the half-value of the maximal initial energy density $\epsilon_{0}(\tau_0)$,
\begin{equation}
S_\mathrm{T}=\pi a b,
\end{equation}
where $a$ and $b$ are the semi-axes of the ellipse.
The $V_{\mathrm{int}}$ values were also divided by the corresponding mean charged particle multiplicities
to avoid the effect of still existing small differences in $\langle dN_{\mathrm{ch}}/d\eta \rangle$
between the experimental set-ups.

\begin{figure}
\centering
\includegraphics[bb=0 0 567 407, width=0.88\textwidth]{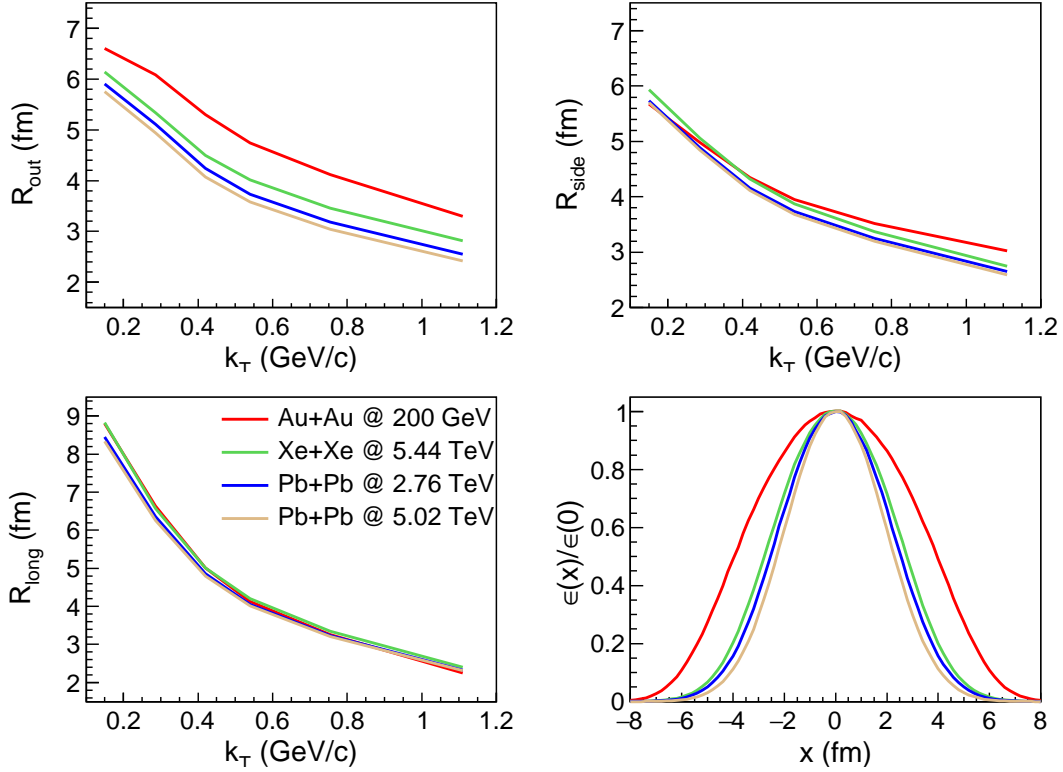}
\caption{The kaon interferometry radii, obtained in iHKM for different collision events
with close charged particle multiplicity (see Table~\ref{tab2} for the details).
The right lower figure demonstrates the corresponding re-scaled initial transverse 
energy-density distributions $\epsilon(x)/\epsilon_0$. 
\label{rk1}} 
\end{figure}

From Fig.~\ref{vint1} it follows that the plotted reduced volume 
\textit{grows} with $S_\mathrm{T}$, having the smallest value for the LHC Pb+Pb collisions at 
$\sqrt{s_{NN}}=5.02$~TeV, a somewhat higher value in the $\sqrt{s_{NN}}=2.76$~TeV case,
after which follows the volume in Xe+Xe collisions at $\sqrt{s_{NN}}=5.44$~TeV,
and the highest one is the value for Au+Au collisions at $\sqrt{s_{NN}}=200$~GeV.
The next plot in Fig.~\ref{rk1} shows the individual kaon interferometry radii dependencies on the pair $k_T$
for the analyzed collision experiments together with the corresponding re-scaled 
initial energy density profiles $\epsilon(x)/\epsilon_0$. 
One can see, that the differences between the $V_{\mathrm{int}}$ values, observed in Fig.~\ref{vint1}, 
are mainly due to the differences in the corresponding transverse radii.
The contribution from $R_{\mathrm{out}}$ radius is the principal one, the $R_{\mathrm{side}}$ contribution
is smaller, and the $R_{\mathrm{long}}$ radii in all the considered cases practically coincide.
The obtained results confirm the assumption, made in~\cite{scaling1,scaling2,pp-lhc}, that
the origin of the observed differences between the interferometry volumes is in different transverse 
flow velocities, which develop in the analyzed systems because of the different initial geometric sizes
and the energy/pressure gradients.

Based on the combined fitting of the longitudinal femtoscopy radii dependency on $k_\mathrm{T}$
and the transverse momentum spectrum, one can estimate the time of maximal emission for particles
of a given species~\cite{mtscale}, i.e. the effective time, that can be associated with the maximal intensity 
of the emission process, when the corresponding particles most actively leave the system.
The presented method was applied by the ALICE Collaboration for the extraction of pion and kaon
times of maximal emission from the experimental data~\cite{alice-scaling}.

The first step is to perform a combined fitting of pion and kaon $p_T$ spectra with the formula 
\begin{equation}
p_0 \frac{d^3N}{d^3p} \propto \exp{[-(m_T/T + \alpha)(1-\bar{v}^2_T)^{1/2}]},
\end{equation}
where $T$ is the effective temperature, $\alpha$ is the parameter, describing the strength of collective flow 
in such a way, that the infinite $\alpha$ means zero flow, and small $\alpha$ means strong flow,
$\bar{v}_T$ is the transverse collective velocity at the saddle point, $\bar{v}_T=k_T/(m_T+\alpha T)$ 
(see Ref.~\cite{mtscale} for details).
From the fit one extracts the common effective temperature of pions and kaons and the two values $\alpha$
for each particle sort.

At the second step one should carry out the fitting of $R_\mathrm{long}(m_T)$ dependencies for pions
and kaons using the formula
\begin{equation}
R^2_{\mathrm{long}}(m_T)=\tau^2\lambda^2\left(1+\frac{3}{2}\lambda^2\right),
\label{rlongfit}
\end{equation}
where $\lambda$ is expressed through the longitudinal homogeneity length of the system $\lambda_l$,  
$\lambda^2=(\lambda_l/\tau)^2=T/m_T\cdot(1-\bar{v}^2_T)^{1/2}$, 
and $\tau$ is the maximal emission time, the one is interested in.
When performing this fit, the $T$ and $\alpha_\pi$ parameters should be constrained within the limits, 
defined earlier during the spectra fitting. However, the $\alpha_K$ parameter for kaons should be left
unconstrained. As a result, one extracts from the fit the desired times of maximal emission $\tau_\pi$ and
$\tau_K$. In the paper~\cite{rhic-ihkm} we obtained the values $\tau_{\pi}=7.12\pm 0.01$~fm/$c$
and $\tau_{K}=9.71 \pm 0.02$~fm/$c$. These results support the hypothesis about the continuous character
of particle emission and freeze-out in high-energy heavy-ion collisions.

\subsection{Direct photon production}
In spite of the hadrons, that are formed at the last stage of superdense system evolution, the direct photon 
production is accumulated from the different sources along with the process
of relativistic heavy ion collision developing. Those include the primary hard photons from the
parton collisions at the very early stage of the process, the photons generated at the pre-thermal
phase of dense matter evolution, then thermal photons at partially equilibrated hydrodynamic
quark-gluon stage, and, finally, from the hadron
gas phase. All the phases of evolution are presented in iHKM, therefore, the corresponding calculations with initial 
conditions providing a good description all the bulk hadron observables (see previous discussions) were done in the 
model for top RHIC~\cite{photons2} and LHC~\cite{photons1} energies. 

\begin{figure}
\centering
\includegraphics[bb=0 0 567 411, width=0.88\textwidth]{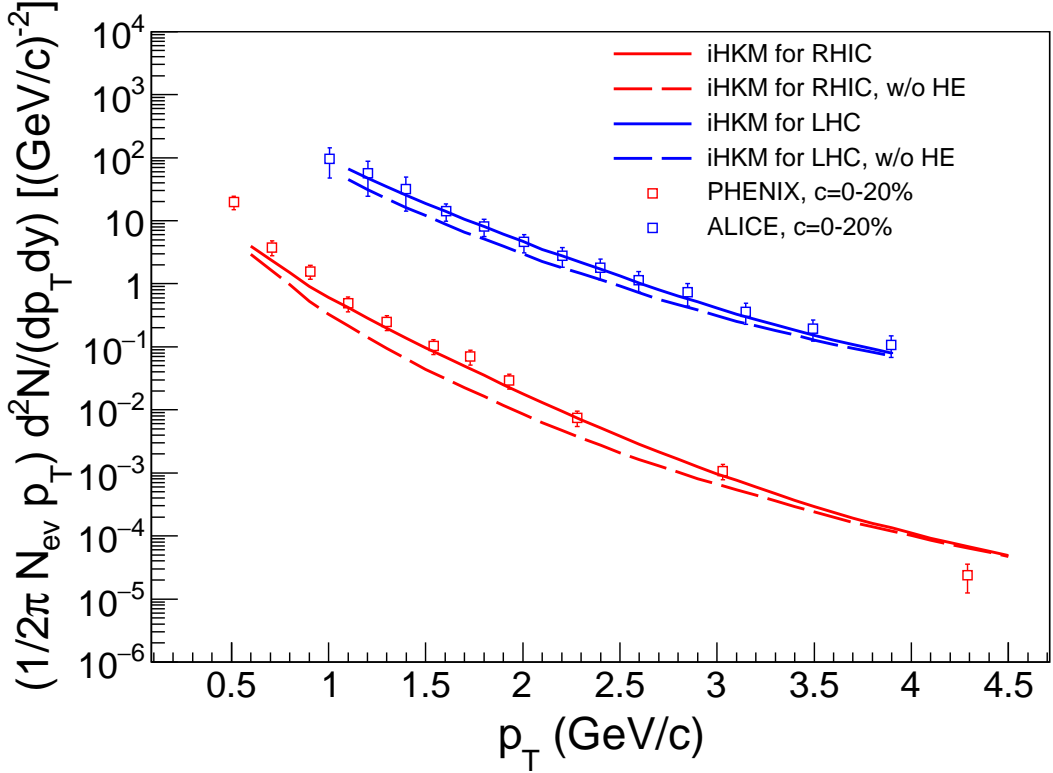}
\caption{The iHKM results for photon spectra in the top RHIC energy and the LHC $2.76 A$~TeV nucleus-nucleus
collisions ($c=0-20\%$) compared to the experimental results~\cite{phenixgam1,alicegam}. The model curves
corresponding to the calculations with and without hadronization emission component are shown.
\label{specgam}} 
\end{figure}

\begin{figure}
\centering
\includegraphics[bb=0 0 567 411, width=0.48\textwidth]{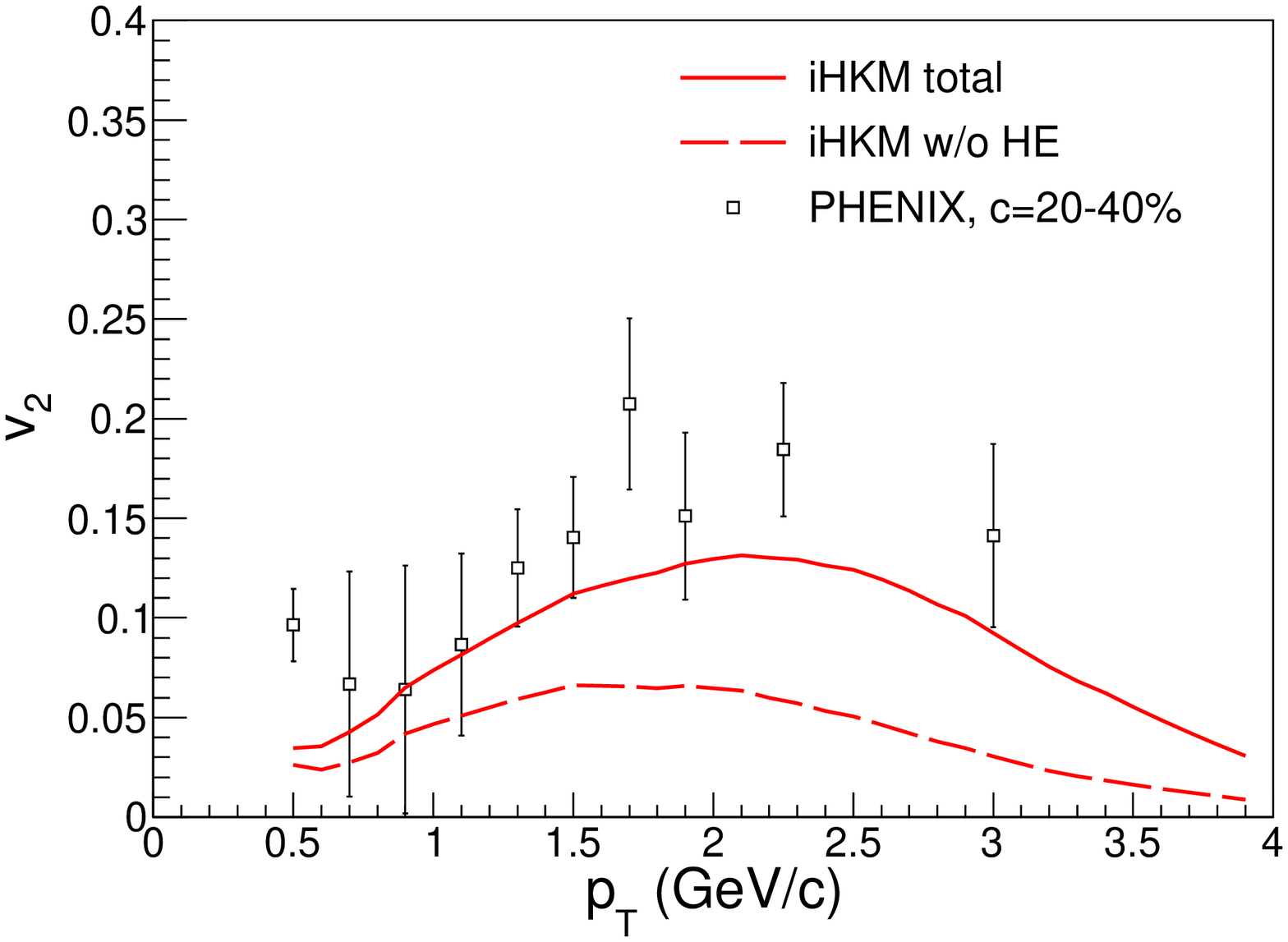}
\includegraphics[bb=0 0 567 411, width=0.48\textwidth]{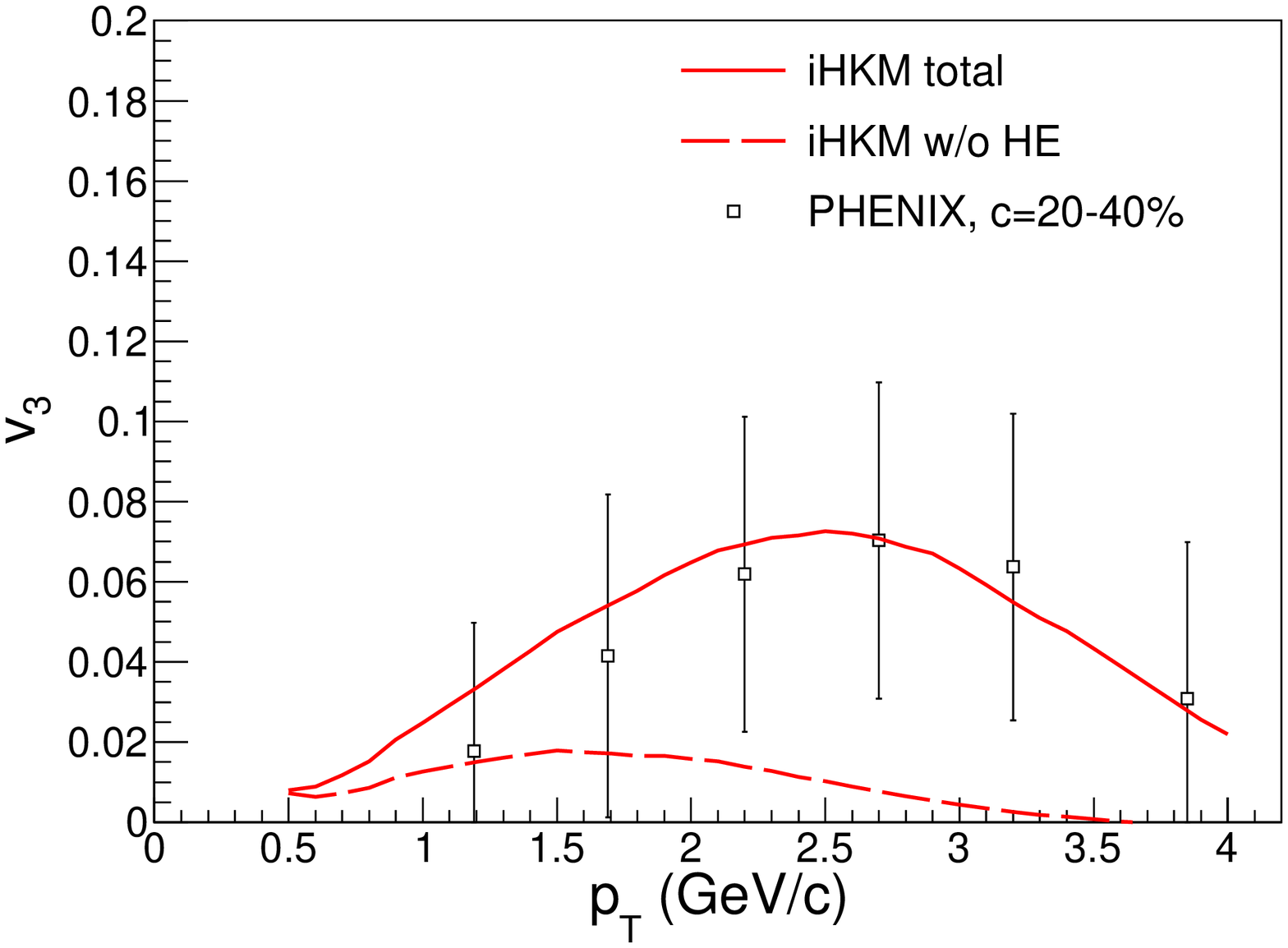}
\caption{The iHKM results for the photon $v_2$ and $v_3$ flow harmonics in the top RHIC energy Au+Au 
collisions ($c=20-40\%$) compared to the experimental results~\cite{phenixgam2}. The model curves
corresponding to the calculations with and without hadronization emission component are shown.
\label{v2gam}} 
\end{figure}

Along the way a hadronic medium evolution is treated in two distinct, in a sense
opposite, approaches: chemically equilibrium and chemically non-equilibrium, namely, chemically
frozen expansion. We find the description of direct photon spectra, elliptic and triangular flow are
significantly improved, for both RHIC the LHC energies, if an additional portion
of photon radiation associated with the confinement processes, the ``hadronization photons'', is
included into consideration. In iHKM this contribution is introduced with only one free parameter, characterizing 
intensity of the photon radiation at the confinement process. This parameter is fixed from fitting the direct photon 
spectra in most central events, and used then at all centralities for spectra, as well as for description of elliptic 
and triangular photon flow. Figure~\ref{specgam} demonstrates the iHKM results for RHIC and LHC photon spectra
in central events ($c=0-20\%$) with and without this hadronization emission (HE). 
The iHKM results for RHIC (centrality 20-40\%) photon elliptic and triangular flows are demonstrated 
in Fig.~\ref{v2gam}. 

The problem of serious underestimation of direct photon production, as well as their flows in theoretical models as 
compared to the experimental data is called ``direct photon puzzle''. From our results it can be concluded, 
that photon radiation in the pseudo-critical region is, probably, a key to solve the photon puzzle.  

\section{Conclusions}
The integrated hydrokinetic model, which simulates the full process of the matter evolution
in course of the relativistic heavy-ion collision, including early pre-equilibrium relaxation dynamics,
that transforms the initially non-equilibrated state of the system to a nearly thermalized one,
allows one to obtain a comprehensive description of measured bulk observables and to reveal
underlying dynamics of the collision process and its femtoscopic space-time structure.
The early start of the system's evolution in iHKM (at proper time about $\tau_0=0.1$~fm/$c$)
allows to account for pre-thermal development of the collective effects and thus improve
the data description.

The model provides a simultaneous description of particle yields and their ratios,
hadron and photon $p_T$ spectra and flow harmonics. Also a good description and succesfull prediction of  pion and kaon femtoscopy radii for all the modern ultrarelativistic 
heavy-ion collision experiments (Au+Au collisions at the RHIC energy $200 A$~GeV,
Pb+Pb collisions at the LHC energies $2.76 A$~TeV and $5.02 A$~TeV, Xe+Xe collisions
at the LHC energy $5.44 A$~TeV) using a single parameter set in each case.
To re-calibrate iHKM, switching from the simulation of one of the mentioned collision types 
to another, one basically needs to change only two parameters, $\alpha$ and $\epsilon_0(\tau_0)$,
and to go from the LHC energy $2.76 A$~TeV to $5.02 A$~TeV --- only one parameter, $\epsilon_0(\tau_0)$.

The simulations carried out using the two different equations of state for quark-gluon phase
with the corresponding particlization temperatures demonstrate, that an equally good description
of data can be achieved in both cases, provided the initial maximal energy density
parameter $\epsilon_0(\tau_0)$ is retuned. This result together with the results of probing 
the hadronic medium with $K^{*}(892)$ resonance, showing that intensive interactions between
hadrons take place at least during 5~fm/$c$ after particlization, speak in favor of continuous
character of chemical and kinetic freeze-outs and indicate a great role of the afterburner stage of 
heavy ion collision in the formation of measured observables.

The iHKM results for the interferometry volumes and radii, obtained for different colliding nuclei
in events with close charged particle multiplicities, allow to conclude, that the hypothesis
about scaling of the femtoscopy radii with multiplicity is not confirmed. The transverse interferometry
radii, and hence, the interferometry volume, depend also on geometric sizes of the colliding nuclei
and grow with the transverse area of their overlapping right after collision.

The investigation of the ``direct photon puzzle'' within iHKM demonstrates that account for the photon radiation 
during the confinement process in the pseudo-critical region can help to solve the problem and provide a good 
description of the direct photon spectra and flow harmonics.

\begin{acknowledgments}
The research was carried out within the project ``Spatiotemporal dynamics and properties of superdense matter in relativistic collisions of nuclei, and their signatures in current experiments at the LHC, RHIC and planned FAIR, NICA''. Agreement № 7/2020 with NAS of Ukraine. It is partially supported by COST Action THOR (CA15213).  
 
\end{acknowledgments}

\end{document}